\documentclass{article}
\pdfoutput=1
\usepackage{graphicx}

\usepackage[utf8]{inputenc}

\usepackage[T1]{fontenc} 
\usepackage{subfigure}
\usepackage{tensor}
\usepackage{amsmath}
\usepackage{amsthm}
\usepackage{amssymb}
\usepackage{color}
\usepackage{hyperref}
\usepackage{fullpage}

\theoremstyle{plain}

\theoremstyle{definition}
\newtheorem{definition}{Definition}[section]

\theoremstyle{remark}
\newtheorem*{remark}{Remark}
\newtheorem{example}{Example}

\numberwithin{equation}{section}

\newcommand{\be}{\begin{equation}}
\newcommand{\ee}{\end{equation}}
\newcommand{\dd}{\ensuremath{\mathrm{d}}} 
\newcommand{\MM}{\ensuremath{\mathcal{M}}} 
\newcommand{\sphere}[1]{\ensuremath{\mathbb{S}^{#1}}} 
\newcommand{\abs}[1] {\ensuremath{\lvert #1 \rvert}} 
\newcommand{\reals}{\ensuremath{\mathbb{R}}} 

\newcommand{\rtilde}{\tilde{r}}

\begin{document}

\title{\bf Graviton time delay and a speed limit for small black holes in Einstein-Gauss-Bonnet theory}

\author{Giuseppe Papallo\footnote{\href{mailto:g.papallo@damtp.cam.ac.uk}{\nolinkurl{g.papallo@damtp.cam.ac.uk}}}\, and Harvey S. Reall\footnote{\href{mailto:h.s.reall@damtp.cam.ac.uk}{\nolinkurl{h.s.reall@damtp.cam.ac.uk}}} \\
\\
\it DAMTP, Centre for Mathematical Sciences, University of Cambridge, \\
\it Wilberforce Road, Cambridge CB3 0WA, UK}

\date{\today}

\maketitle
\begin{abstract}
Camanho, Edelstein, Maldacena and Zhiboedov have shown that gravitons can experience a negative Shapiro time delay, i.e. a time advance, in Einstein-Gauss-Bonnet theory. They studied gravitons propagating in singular ``shock-wave'' geometries. We study this effect for gravitons propagating in smooth black hole spacetimes. For a small enough black hole, we find that gravitons of appropriate polarisation, and small impact parameter, can experience time advance. Such gravitons can also exhibit a deflection angle less than $\pi$, characteristic of a repulsive short-distance gravitational interaction. We discuss problems with the suggestion that the time advance can be used to build a ``time machine''. In particular, we argue that a small black hole cannot be boosted to a speed arbitrarily close to the speed of light, as would be required in such a construction. 
\end{abstract}

\section{Introduction} 
\label{sec:introduction}

Lovelock theories of gravity \cite{Lovelock1971} are an interesting alternative to General Relativity in $d>4$ dimensions. It is interesting to ask how classical properties of these theories differ from GR, in particular whether these theories are pathological in some way. 
 
In GR, causality is determined by the null cone of the metric. It has been known for a long time that this is not the case in Lovelock theories. In general, causal properties of a PDE are determined by its {\it characteristic hypersurfaces} \cite{CourantHilbert}. For example: sharp features of solutions (e.g. a discontinuity in curvature or in the 100th derivative of curvature) propagate along such surfaces; for high-frequency waves propagating in a background spacetime, the surfaces of constant phase are characteristic hypersurfaces of the background; given some initial data surface $\Sigma$, and a region $\Omega \subset \Sigma$, the region of spacetime where the solution is determined by data specified on $\Omega$ is bounded by an ingoing characteristic surface emanating from $\partial \Omega$.

In GR, a hypersurface is characteristic if, and only if, it is null w.r.t. the metric. However, in Lovelock theories, the location of characteristic hypersurfaces depends on both the metric and the Riemann tensor. Generically, such surfaces are non-null w.r.t. the metric: gravity can propagate faster, or slower, than light \cite{Aragone1988,Choquet-Bruhat1988}. Furthermore, different graviton polarizations generically propagate along different characteristic hypersurfaces (i.e. at different ``speeds'') so such theories are ``multirefringent''. 
  
Another important difference between GR and Lovelock theories has been demonstrated recently \cite{Reall2014}. In GR, the equation of motion is always hyperbolic: it has the character of a nonlinear wave equation propagating $d(d-3)/2$ degrees of freedom in $d$ spacetime dimensions. For small curvature, the same is expected to be true in Lovelock theories (although this has not been proved). However, the equation of motion can be non-hyperbolic when the curvature is large. Under such circumstances, the initial value problem is no longer well-posed. Consider initial data, consisting (as in GR) of  a $(d-1)$-manifold $\Sigma$, a Riemannian metric $h_{ab}$ on $\Sigma$ and a symmetric tensor $K_{ab}$ (the extrinsic curvature) on $\Sigma$. From this data one can determine whether or not the equation of motion is hyperbolic on $\Sigma$. If it is not hyperbolic then, generically, there will exist no solution of the equation of motion arising from such initial data: one cannot ``evolve forward in time''. Solutions may exist in non-generic situations e.g. for analytic initial data the Cauchy-Kowalevskaya theorem may guarantee local existence of a solution. However, this is infinitely fine-tuned: if one smoothly perturbs the initial data within a compact region then generically there will exist no solution of the equation of motion arising from the perturbed data. 

As an example, consider Einstein-Gauss-Bonnet (EGB) theory (a Lovelock theory). This theory admits static, spherically symmetric black hole solutions  \cite{Boulware1985}. Since EGB theory has a dimensionful coupling constant, we can talk about ``small'' and ``large'' black holes. Ref.~\cite{Reall2014} showed (using results of Refs.~\cite{Dotti2005a,Gleiser2005}) that, for a small black hole with $d=5$ or $6$, the equation of motion is non-hyperbolic in a region outside the event horizon. Hence such a black hole solution must be rejected as unphysical: it is not the Cauchy evolution of initial data for which the equation of motion is hyperbolic.\footnote{Of course, one could regard such a black hole solution as the evolution of initial data for which the equation of motion is not everywhere hyperbolic e.g. a surface of constant $t$. However, as just explained, the violation of hyperbolicity implies that this is infinitely fine-tuned: if one perturbs such initial data then, generically, it will be impossible to evolve the perturbed data.}

As we explain in the Appendix, a hyperbolic PDE defines a natural division of vectors into ``timelike'', ``spacelike'' and ``null'' such that causal propagation occurs along non-spacelike directions. In GR, these definitions coincide with the notions of ``timelike'' etc defined using the metric but this is not the case in Lovelock theories. Since the causal structure defined by the equation of motion is more important than that defined by the metric, we will adopt the following convention:

\medskip

 {\it Throughout this paper, the words ``timelike'', ``spacelike'', ``null'' or ``causal'' refer to the causal structure defined by the equation of motion, {\it not} to the causal structure defined by the metric. If we wish to refer to the latter then we will say ``timelike w.r.t. the metric'' etc.}

\medskip

There is another restriction on initial data which needs to be mentioned. Given a hyperbolic PDE, for a well-posed Cauchy problem, initial data must be chosen so that the initial data surface $\Sigma$ is spacelike (in the sense just defined). In GR, this is the same as being spacelike w.r.t. the metric but this is not the case in Lovelock theories. One can find examples of initial data for which $\Sigma$ is everywhere spacelike w.r.t. the metric but not everywhere spacelike. Evolution of such initial data is not a well-posed problem. We will discuss this further below. 

In summary, there are several restrictions on legitimate initial data in Lovelock theories: data must be chosen so that (i) the constraint equations are satisfied; (ii) the equation of motion is hyperbolic; and (iii) the initial data surface is spacelike. 

The present paper is motivated by recent work of Camanho, Edelstein, Maldacena and Zhiboedov (CEMZ) \cite{Camanho2014}. They investigated the Shapiro time delay for gravitons in EGB theory. In this theory, linearised gravity propagates at the speed of light in a flat background but can travel faster than light in a curved background. Hence one might expect that it is possible for a gravitational signal to travel through a curved spacetime ``faster'' than it would in flat spacetime. CEMZ showed that this is indeed the case. They considered the Aichelburg-Sexl (AS) ``shock-wave'' spacetime \cite{Aichelburg1971,Dray1985}. This is an exact solution of EGB theory which CEMZ interpreted as describing the gravitational field of a high-energy particle e.g. a graviton. They showed that a second (``test'') graviton that scatters from the first graviton at small impact parameter can indeed experience a negative Shapiro time delay, i.e., a time advance. They also showed that the same result can be obtained from a scattering amplitude calculation.

The AS solution is singular: it has delta function curvature localised on a null hypersurface, with the amplitude of the delta function diverging along a line within this hypersurface (the worldline of the high-energy graviton). Given that the solution is singular, one must ask whether it is physical. The usual reason why the AS solution is regarded as physical is because it can be obtained as a limit of smooth black hole solutions.  The limit involves boosting a black hole and taking the boost to infinity whilst scaling the mass to zero, keeping the energy fixed. Hence one can ``regulate'' a singular AS solution by replacing it with a small, highly boosted, black hole. 

This motivation for regarding the AS solution as physical is fine in GR but it suffers problems in EGB theory. First, for $d=5,6$, small black hole solutions are unphysical, as explained above.\footnote{In fact, for $d=5$, there is a mass gap so a black hole cannot have arbitrarily small mass.} There is also a problem for $d \ge 7$. Consider initial data describing a boosted black hole. We will show that, for a small black hole, if the boost is too large then such initial data is not everywhere spacelike (in the sense defined above). This implies that if one makes a generic smooth perturbation to this initial data then there will not exist a solution to the equation of motion arising from the perturbed data. Therefore such initial data is unphysical: it cannot arise in Cauchy evolution without infinite fine-tuning. 
Hence, in EGB theory, a small black hole cannot be boosted arbitrarily close to the speed of light: there is a speed limit. This suggests that the AS solution may not be physical in EGB theory. Therefore it seems desirable to have an independent confirmation that time advance is possible. 

We will study the Shapiro time delay for gravitons in static, spherically symmetric, black hole solutions of EGB theory. Another motivation for doing this is that, in general, there is no gauge-invariant definition of the Shapiro time delay \cite{Gao2000}. The problem is that there is no gauge-invariant way of identifying points of a curved spacetime with those of Minkowski spacetime. However, this problem can be overcome for spacetimes with suitable symmetries. In particular, for static, spherically symmetric spacetimes, it {\it is} possible to define the Shapiro time delay unambiguously \cite{CabreraPalmer2003}.\footnote{The large symmetry of the AS solution implies that the time delay should be unambiguous in this case too.} The idea is to consider a spherical cavity surrounding the black hole and calculate the (proper) time it takes for a graviton to cross the cavity, and then compare with the time it takes to travel between the corresponding points of a spherical cavity in flat spacetime. 

Our main result is to confirm that a time advance for gravitons is indeed possible for small black hole solutions of EGB theory. This occurs for gravitons encountering the black hole with appropriate polarisation, and an impact parameter comparable to the length scale set by the GB coupling, just as CEMZ found for the AS spacetime. Furthermore, we find that such gravitons can undergo a deflection through an angle less than $\pi$. Both of these features indicate that certain graviton polarizations experience a {\it repulsive} gravitational interaction at distances comparable to the scale set by the coupling constant, which is confirmed by examining the ``effective potential'' which determines graviton trajectories. Close to the black hole horizon, gravity becomes attractive again, and the deflection angle is greater than $\pi$ for very small impact parameter. 

Time advance can occur when the cavity is not too large. For a very large cavity, the time delay is positive. This is because the graviton suffers a deflection, which makes its path longer than the corresponding (straight line) path in flat spacetime.\footnote{CEMZ evaded this effect by considering a graviton propagating between two AS shocks, resulting in zero deflection.} This results in a contribution to the time delay which grows in proportion to the cavity radius, ensuring that the delay is positive for large enough cavity radius. However, there are two special values for the impact parameter for which the deflection angle is exactly $\pi$, i.e., the graviton suffers no net deflection. For these, the time delay is finite in the limit of infinite cavity radius. For the larger of these values, the time delay is positive. However, for the smaller value it can be negative. So this is an example of time advance for a graviton incident from, and returning to, infinity. 

CEMZ argued that the possibility of a time advance implies that EGB is unviable as a classical theory. They employed an argument which asserts that when one has superluminal propagation in a Lorentz covariant theory one can construct a time machine, i.e., a spacetime with closed causal curves (where ``causal'' is defined as above). Arguments of this sort have been applied to various non-gravitational field theories in Ref.~\cite{Adams2006}. However, this kind of argument has been criticised by Geroch \cite{Geroch2010}, who notes that the existence of a causally pathological solution is not enough to reject the theory (see also Ref. \cite{Bruneton2007}). After all, GR admits causally pathological solutions e.g. Minkowski spacetime with a periodic time direction. The existence of such solutions is not regarded as problematic because one cannot ``form'' them, i.e., they do not arise as the Cauchy evolution of initial data. We will argue that the ``time machine'' constructions of Ref.~\cite{Adams2006} and CEMZ also do not arise as the Cauchy evolution of legitimate initial data. The reason is that the  initial data surface is not everywhere spacelike (in the sense defined above). Hence it is not a well-posed problem to specify a solution in terms of such data: either no solution exists, or it is infinitely fine-tuned.

If the ``time machine'' construction doesn't work then is there some other reason to regard EGB theory, or more general Lovelock theories,  as ``badly behaved'' when compared to GR? There is no proof that the initial value problem for Lovelock theories is well-posed, even for initial data for which the equation of motion is hyperbolic and the initial surface is spacelike e.g. small curvature initial data.\footnote{See Ref. \cite{Willison2014,Willison2015} for discussion of the first steps in constructing such a proof.} If the initial value problem is not well-posed then these theories don't make sense classically. Assuming that the initial value problem {\it is} well-posed, Ref.~\cite{Reall2014a} argued that Lovelock theories, unlike GR, will suffer from shock formation similar to a compressible perfect fluid. Since different parts of a wavepacket can travel with different speeds, one could construct a smooth wavepacket for which the back travels faster than the front, leading to wave steepening and, eventually, a shock. The curvature would diverge at a shock. Ref.~\cite{Reall2014a} argued that this would be a naked singularity. So this is an example of a pathology that afflicts Lovelock theories but not GR. However, it was argued to be a ``large data'' effect, i.e., it may not occur for small curvature initial data. Indeed, Ref.~\cite{Reall2014a} argued that Minkowski spacetime is likely to be nonlinearly stable in Lovelock theories.

This paper is organised as follows. We first give an informal review of the notions of hyperbolicity and causality required in EGB theory (Section~\ref{sec:causal_structure}). In Section~\ref{sec:EGB_black_holes} we review the static, spherically symmetric, black hole solutions of EGB and derive a ``speed limit'' for small black holes. In Section~\ref{sec:time_delay} we will investigate the Shapiro time delay and deflection of gravitons propagating in a static, spherically symmetric, black holes solution of EGB. In Section~\ref{sec:time_machines} we discuss the ``time machine'' constructions of Ref.~\cite{Adams2006} and CEMZ. Appendix~\ref{sec:causal_structure_appendix} gives a more technical review of how a hyperbolic PDE defines a notion of causal structure. Appendix~\ref{sec:approximation} gives details of our perturbative calculation of the time delay and deflection angle. 

\section{Hyperbolicity and causal structure} 
\label{sec:causal_structure}

In this section we will give a brief review of the definitions of hyperbolicity and causality that we will need. A more technical review is presented in Appendix~\ref{sec:causal_structure_appendix}. Hyperbolic PDEs describe phenomena that exhibit finite propagation speed. A hyperbolic PDE defines a causal structure: given a spacetime point $p$, one can define the causal past of $p$ as the set of points $q$ for which a change in the solution at $q$ would result in a change in the solution at $p$. Similarly the causal future of $p$ is the set of points $q$ where the solution at $q$ which depend on the solution at $p$. In general, the causal past/future depend on the solution itself.

In general, the causal structure is determined by the coefficients of the highest order derivatives in the PDE. 
The causal structure of a PDE is closely related to its \emph{characteristic hypersurfaces}. Loosely speaking these are hypersurfaces on which all of the highest order derivatives cannot be determined in terms of all the lower derivatives by the equations of motion. If one has a solution which is discontinuous in the highest derivatives (but continuous in the others) across a hypersurface, then that hypersurface is necessarily characteristic. Hence, discontinuities in the solution propagate along characteristics. More ``physically'', if one considers high frequency waves propagating on top of a ``background'' solution then the surfaces of constant phase are characteristic hypersurfaces of the background solution. 

There is an algebraic criterion to determine whether or not a hypersurface is characteristic. Given a point $p$ and a covector $\xi$ at $p$ one can define the 
\emph{characteristic polynomial} $Q(p,\xi)$, a homogeneous polynomial in $\xi$ with coefficients determined by the coefficients of the highest derivative terms in the PDE. A hypersurface \(\Sigma\) is characteristic if, and only if, its normal one-form \(\xi\) satisfies:
\begin{equation}
	Q(p,\xi)=0
\end{equation}
\noindent for any \(p\in\Sigma\). At each point, the set of one-forms satisfying this equation defines a cone in the cotangent space, known as \emph{normal cone}. 

In GR, the normal cone is simply the light cone in the cotangent space. In Lovelock theories, the normal cone is generically non-null w.r.t. the metric \cite{Aragone1988,Choquet-Bruhat1988}. In GR, the $d(d-3)/2$ degrees of freedom of the metric all propagate with the same speed. However, in Lovelock theories these degrees of freedom generically propagate at different speeds so the normal cone is expected to be composed, generically, of $d(d-3)/2$ distinct ``sheets''.\footnote{This has been proved only for the class of Ricci flat type N spacetimes \cite{Reall2014}.} For linearised perturbations of a solution, this corresponds to the statements that different graviton polarizations propagate with different speeds: Lovelock theories are \emph{multirefringent}.

The definition of hyperbolicity is reviewed in the Appendix. Roughly speaking, this amounts to requiring the ``correct number'' of propagating degrees of freedom at each point. For the equation of motion to be hyperbolic, we require that the normal cone has $d(d-3)/2$ sheets. Obviously this isn't true in GR, so one has to allow for degeneracy i.e. the possibility that multiple graviton polarizations propagate along the same sheet. We also require that the interiors of these sheets have a non-trivial intersection. We then define a covector to be timelike if it lies in the interior of this intersection, causal if it lies in the closure of this intersection (and is non-zero), and spacelike if it lies outside this closure. A hypersurface is said to be spacelike if its normal covector is timelike everywhere. Hence if the theory is hyperbolic then (i) we have the right number of degrees of freedom and (ii) we can find surfaces that are ``spacelike for all degrees of freedom''. 

The dual of the normal cone, is another cone, this time in the tangent space, known as the \emph{ray cone}. In the case of the linear wave equation or General Relativity this is simply the null cone defined by the spacetime metric. In a Lovelock theory it will generically be composed of $d(d-3)/2$ distinct sheets. The ray cone is generated by {\it bicharacteristic curves}. These are determined, in local coordinates $x^\mu$, by solving the Hamiltonian equations
\be
 \frac{d x^\mu}{ds} = \frac{\partial Q}{\partial \xi_\mu} \qquad \frac{d \xi_\mu}{ds} = -\frac{\partial Q}{\partial x^\mu}
\ee
with $\xi(0)$ chose so that $Q(x(0),\xi(0))=0$. For the wave equation, or for GR, bicharacteristic curves correspond to null geodesics. For high frequency waves, one can use the geometric optics approximation to show that high-frequency gravitons follow null geodesics. Similarly, in a Lovelock theory, high-frequency graviton worldlines are bicharacteristic curves (see e.g. \cite{Reall2014a}) so we will calculate the Shapiro time delay for such gravitons by determining these curves.  

We define the \emph{causal cone} as the convex hull of the sheets defining the ray cone. We will then define a vector to be \emph{causal} if it lies in its closure and \emph{spacelike} if it lies in its exterior. Note that a hypersurface  is spacelike iff all tangent vectors are spacelike.

To illustrate the above definitions, consider the following example. For some non-generic spacetimes (possessing appropriate symmetries) the characteristic polynomial factorizes into a product of quadratic factors:
\begin{equation}
	Q(p,\xi)=(G_{1}(p)^{ab}\xi_{a}\xi_{b})^{q_{1}}(G_{2}(p)^{ab}\xi_{a}\xi_{b})^{q_{2}}\dots
\end{equation}
For example, this happens for any Ricci flat spacetime with a type N Weyl tensor, and also for the static spherically symmetric spacetimes that we will consider below  \cite{Reall2014}. Each of the \(G_{I}^{ab}\) can be interpreted as an (inverse) \emph{effective metric}. A hypersurface is characteristic iff its normal is null w.r.t. one of these effective metrics. The normal cone is simply the union of the null cones of $G_I^{ab}$. In this case, the theory is hyperbolic iff (i) there are the ``right number'' of effective metrics, i.e., one for each degree of freedom (allowing for the possibility that one effective metric may correspond to several degrees of freedom); (ii) each $G_I^{ab}$ has Lorentzian signature; and (iii) the interiors of the null cones of $G_I^{ab}$ have a non-trivial intersection. This is the case for any Ricci flat type N spacetime \cite{Reall2014}. However, as we will see below, for certain {\it small} black hole solutions, it turns out that one of the $G_I^{ab}$ can fail to be everywhere Lorentzian, so the theory is not hyperbolic in such spacetimes \cite{Reall2014}. For both examples, when the theory is hyperbolic, it turns out that the null cones of $G_I^{ab}$ form a nested set, so causality (for covectors) is determined by the effective metric corresponding to the innermost null cone. We define $G_{Iab}$ to be the inverse of $G_I^{ab}$. Then the ray cone is the union of the null cones of $G_{Iab}$ and causality (for vectors) is determined by the effective metric with the outermost null cone. The bicharacteristic curves are simply the null geodesics of the $G_{Iab}$. 


\section{Einstein-Gauss-Bonnet black holes}
\label{sec:EGB_black_holes}
\subsection{Spherically symmetric EGB black holes} 
\label{sub:spherically_symmetric_egb_black_holes}

The equation of motion of Einstein-Gauss-Bonnet (EGB) theory is obtained by varying the action 
\be
 S =\frac{1}{16\pi} \int d^d x \sqrt{-g} \left[ R + \lambda_{\rm GB}(R^{2}-4 R_{ab}R^{ab}+R_{abcd}R^{abcd})\right]
\ee
where $\lambda_{GB}$ is the GB coupling constant and we have set $G=1$. This theory admits static, spherically symmetric solutions with metric \cite{Boulware1985}:
\begin{equation}
	g=-f(r)\dd t^2 + f(r)^{-1} \dd r^2 +r^2 g_{\sphere{d-2}}
	\label{eq:sph_sym_metric}
\end{equation}
\noindent where \(g_{\sphere{d-2}}\) is the standard round metric on the unit (\(d-2\))-sphere \(\sphere{d-2}\) and
\begin{align}
	\label{eq:f_EGB}
	f(r)&=1+\frac{r^2}{\alpha}(1-q(r))\\
	\label{eq:q_EGB}
	q(r)&=\sqrt{1+\frac{2\alpha\mu}{r^{d-1}}}
\end{align}
\be
 \alpha = 2 (d-3) (d-4) \lambda_{\rm GB}
\ee
and we assume $\alpha>0$ since for \(\alpha<0\) the metric has a naked singularity \cite{Boulware1985}.\footnote{A priori, the sign in front of \(q\) in the expression for \(f\) is arbitrary. We choose the negative branch, corresponding to asymptotically flat solutions.} The parameter \(\mu\) is proportional to the ADM mass $M$:
\be
M=\frac{(d-2)\mathrm{Vol}(\sphere{d-2})}{16\pi}\mu.
\ee
The event horizon is a hypersurface \(r=r_{H}\),\footnote{Since some modes of gravitational perturbations can travel faster than light, it could be possible, a priori, that such perturbations could escape the black hole region (as defined by the causal structure given by the physical metric). However it was shown in \cite{Izumi2014,Reall2014} that a Killing horizon is always a characteristic hypersurface for all graviton polarizations, excluding this possibility when the event horizon is a Killing horizon, as is the case here.} where $r_H$ is the largest root of
\begin{equation}
\label{rHdef}
	\mu=r_{H}^{d-5}\left(r_{H}^2+\frac{\alpha}{2}\right).
\end{equation}
The coupling $\alpha$ has dimensions of length squared so EGB theory has a length scale \(\sqrt{\alpha}\). We will say that a black hole is ``small'' if $r_H \ll \sqrt{\alpha}$ and ``large'' if $r_H \gg \sqrt{\alpha}$. Equivalently, for \(d>5\), a black hole is small if 
\be
\label{smallbh}
 \mu \ll \alpha^{(d-3)/2}
\ee
and large if $\mu \gg \alpha^{(d-3)/2}$. Note that black holes with arbitrarily small mass do not exist for $d=5$ because there is a mass gap (however \(r_{H}\) can be arbitrarily small): $\mu > \alpha/2$.  
The function $q$ varies over a length scale
\be
\label{Ldef}
L \equiv (\mu\alpha)^{\frac{1}{d-1}}.
\ee
For $r \gg L$ we have
\be
\label{fexp}
 f \approx 1 - \frac{\mu}{r^{d-3}}
\ee
i.e., the solution reduces to the higher-dimensional Schwarzschild solution. For a large black hole $r_H \gg L$ so this approximation is valid everywhere outside the horizon. 

\subsection{Effective metrics and bicharacteristic curves}

Characteristic hypersurfaces of the above solution were determined in Ref. \cite{Reall2014}. The symmetries of the solution imply that the characteristic polynomial factorizes into a product of quadratic factors, each associated to an ``effective metric'' as discussed above. A hypersurface is characteristic iff it is null w.r.t. one of these effective metrics. The explicit form of the effective metrics was determined by considering linear perturbations of such solutions. Such perturbations can be classified into scalar (S), vector (V) and tensor (T) types w.r.t. the spherical symmetry. For each type, one can obtain a ``master equation'' ~\cite{Dotti2005a,Gleiser2005} and from these one can read off the effective metric for that type. Hence the characteristic polynomial factorizes as
\begin{equation}
	Q(p,\xi)=(G_{S}^{ab}\xi_a \xi_b)^{p_{S}}(G_{V}^{cd}\xi_c \xi_d)^{p_{V}}(G_{T}^{ef} \xi_e \xi_f)^{p_{T}}
\end{equation}
\noindent where $p_S, p_V, p_T$ denote the number of degrees of freedom of each type of modes. Viewing $G_S^{ab}$ etc as inverse metrics, the corresponding metrics are given by
\begin{equation}
	\label{eq:effective_metrics}
	G_{A}=-f(r)\dd t^2 + f(r)^{-1} \dd r^2 +\frac{r^2}{c_{A}(r)} g_{\sphere{d-2}}
\end{equation}
for certain smooth functions $c_A(r)$ given by \cite{Reall2014}
\begin{align}
	\label{eq:CS_EGB}
	c_{S}(r)&=3\left(1-\frac{1}{d-2}\right){\cal A}(r)+\left(1-\frac{3}{d-2}\right)\frac{1}{{\cal A}(r)}-3\left(1-\frac{2}{d-2}\right),\\
	\label{eq:CV_EGB}
	c_{V}(r)&={\cal A}(r),\\
	\label{eq:CT_EGB}
	c_{T}(r)&=-\left(1+\frac{1}{d-4}\right){\cal A}(r)-\left(1-\frac{1}{d-4}\right)\frac{1}{{\cal A}(r)}+3.
\end{align}
where
\be
\label{eq:A_EGB}
	{\cal A}(r)=q(r)^{-2}\left(\frac{1}{2}+\frac{1}{d-3}\right)+\left(\frac{1}{2}-\frac{1}{d-3}\right),
\ee
It is convenient to take the index $A \in \{0,S,V,T \}$ where $0$ refers to the physical metric, i.e.,
\be
 c_0(r)  \equiv 1.
\ee
For a large black hole, the functions $c_A(r)$ are positive everywhere outside the horizon. This ensures that the effective metrics have Lorentzian signature, and their null cones form a nested set, with the outermost cone (in the tangent space) corresponding to the effective metric with the largest value of $c_A(r)$. (The physical metric $G_0$ is included in this nested set.) This ensures the hyperbolicity of the theory, and causality is determined by this outermost null cone. If $c_A>1$ then the associated modes propagate faster than light \cite{Reall2014}.

For $d=5,6$, for small enough $r_H$, it turns out that one of the $c_A(r)$ vanishes at some value $r=r_*>r_H$ and becomes negative for $r<r_*$ \cite{Reall2014}. The corresponding inverse effective metric $G_A^{ab}$ is smooth at $r=r_*$ but becomes degenerate there. For $r<r_*$ it has Lorentzian signature, but with the opposite overall sign. This implies that the theory is non-hyperbolic in such black hole spacetimes. Therefore small black holes are unphysical for $d=5,6$. This does not occur for EGB theory with $d \ge 7$.\footnote{However, it does occur for more general Lovelock theories with $d \ge 7$. Generically, it occurs when the equation of motion includes the highest order Lovelock term \cite{Reall2014}.} 

Note that the function $c_A$ are determined entirely by the length scale $L$ defined by (\ref{Ldef}). If $r \gg L$ then
\be
\label{cexp}
 c_A(r) = 1 + 2 \beta_A \left( \frac{L}{r} \right)^{d-1} + {\cal O}\left( \left( \frac{L}{r} \right)^{2(d-1)} \right)
\ee
where the constants $\beta_A$ are given by
\begin{equation}
	\beta_{S}=-\frac{(d-1)}{(d-3)},\quad
	\beta_{V}=-\frac{1}{2}\frac{(d-1)}{(d-3)},\quad
	\beta_{T}=\frac{(d-1)}{(d-3)(d-4)}, \quad 
	\beta_0 = 0.
\end{equation}
We see that $c_S,c_V<1$ at large $r$ so the scalar and vector polarizations propagate slower than light in this region. However $c_T>1$ at large $r$ so tensor polarizations propagate faster than light at large $r$. Hence causality at large $r$ is determined by the effective metric for the tensor modes. 

We will now prove that $c_S<1$ and $c_V<1$ {\it everywhere}. Since \(q(r)^{-2}\) is monotonically increasing  we see that also \({\cal A}(r)\) is monotonically increasing. We also have ${\cal A}(\infty)=1$. It follows that ${\cal A}(r)<1$ hence $c_V(r)<1$. Now we look at \(c_{S}(r)\). We have: 
\begin{equation}
c_{S}'(r)=\left[3\left(1-\frac{1}{d-2}\right)-\left(1-\frac{3}{d-2}\right){\cal A}(r)^{-2}\right]{\cal A}'(r).
\end{equation}
Since \({\cal A}(r)\) is monotonically increasing, the sign is determined by the terms in parentheses. For $d=5$ this is constant and positive, hence $c_S'>0$ so $c_S(r)<c_S(\infty)=1$. For $d \ne 5$, the expression in parentheses is negative at small $r$ and positive at large $r$. Hence, starting from $r=0$, $c_S(r)$ decreases to a minimum and then increases monotonically with $r$. Hence $c_S(r) <\max\{c_{S}(\infty),c_{S}(0)\}=c_{S}(\infty)=1$.

The same argument allows us to determine an upper bound for $c_T$. We have: 
\begin{equation}
c_{T}'(r)=-\left[\left(1+\frac{1}{d-4}\right)-\left(1-\frac{1}{d-4}\right){\cal A}(r)^{-2}\right]{\cal A}'(r).
\end{equation}
If \(d=5\) then the expression in square brackets is constant and positive so we see that \(c_{T}\) is monotonically decreasing hence
\be
 c_T(r) < c_T(0) = 3 \qquad (d=5).
\ee
If \(d> 5\) then $c_T$ has a maximum at \(r_0\) where 
\be
{\cal A}(r_0)=\sqrt{\frac{1-1/(d-4)}{1+1/(d-4)}}
\ee
and hence 
\be
\label{eq:cmax}
c_{T}(r)<c_{T}(r_0)=3-2\sqrt{1-\frac{1}{(d-4)^{2}}}
\ee
Note that the RHS is greater than $1$. 

\subsection{Speed limit for small black holes}
\label{sec:speedlimit}

We will now consider the effect of boosting one of these black holes. To construct initial data describing a boosted black hole, we can consider the data induced on a boosted hypersurface in the black hole spacetime. Such a hypersurface is spacelike w.r.t. the metric for any boost velocity $v$ such that $|v|<1$. However, since the null cone for the tensor modes can lie outside the null cone of the physical metric, it is possible that, for $|v|$ close to $1$, the hypersurface may fail to be everywhere spacelike w.r.t. the tensor effective metric. This implies that it will fail to be spacelike in the sense defined in the Introduction and hence it would not be a valid initial data surface. We will now show that this is indeed what happens.

First we introduce an ``isotropic'' radial coordinate $\tilde{r}$ defined by
\be
\frac{d\log \tilde{r}}{dr} = \frac{1}{r \sqrt{f}}
\ee
so that the physical metric is
\be
g= -f dt^2 + H \left( d\tilde{r}^2 + \tilde{r}^2  g_{\sphere{d-2}}\right)
\ee
where
\be
 H = \frac{r^2}{\tilde{r}^2}.
\ee
For $r \gg L$ we can use the approximation (\ref{fexp}) to obtain
\be
 \rtilde\approx r\left(1-\frac{\mu}{2(d-3) r^{d-3}}\right)
\ee
and hence
\be
 f \approx 1- \frac{\mu}{\tilde{r}^{d-3}} \qquad H \approx 1+\frac{\mu}{(d-3) \tilde{r}^{d-3}}.
\ee 
To construct initial data describing a boosted black hole we convert to Cartesian coordinates $x^i$ so that $x^1 = \tilde{r} \cos \theta_1$ etc (where $\theta_1,\theta_2,\ldots$ are the angles on $S^{d-2}$) and then perform the Lorentz transformation
 \begin{equation}
	x^1 = \gamma (x^{1'}-vt'), \quad t= \gamma (t'-vx^{1'}), \qquad \gamma = (1-v^2)^{-1/2}.
\end{equation}
We now consider the data induced on a surface of constant $t'$. By inverting the Lorentz transformation, we see that this is the same as the data induced on a surface of constant $t+vx^1$, i.e., a surface of constant $t+v\tilde{r} \cos \theta_1$. Let $\Sigma$ be such a surface. Define a 1-form $\xi$ normal to $\Sigma$:
\be
  \xi = dt + v \cos \theta_1 d\tilde{r} - v\tilde{r} \sin \theta_1 d\theta_1.
\ee
We want to take the data induced on $\Sigma$ as initial data describing a boosted black hole. To do this, we must check that $\Sigma$ is spacelike in the sense defined in the Introduction. This is equivalent to $\Sigma$ being spacelike w.r.t. all of the effective metrics, i.e., $\xi$ must be timelike w.r.t. to all of the effective metrics. To investigate whether this is this case, consider the norm of $\xi$ w.r.t. $G_A$ at $\theta_1 = \pi/2$:
\be
 G_A^{\mu\nu} \xi_\mu \xi_\nu |_{\theta_1 = \pi/2} = -f^{-1} + v^2 c_A \frac{\tilde{r}^2}{r^2}.
\ee
Now assume that the black hole is small and consider the region $L \ll r \ll \sqrt{\alpha}$ where $\mu/r^{d-3} \ll L^{d-1} /r^{d-1}$. Using our expansion for $c_A$ we then obtain
\be
 G_A^{\mu\nu} \xi_\mu \xi_\nu |_{\theta_1 = \pi/2}  \approx -(1-v^2) + 2 \beta_A v^2 \left( \frac{L}{r} \right)^{d-1}.
\ee
For the scalar, vector and physical metrics we have $\beta_A \le 0$ so the RHS is always negative. However, for the tensor effective metric we have $\beta_T>0$ and hence if $v$ is too close to $1$ then the second term above, although small, will overwhelm the first and the RHS will be positive, i.e., $\xi$ will be spacelike w.r.t. $G_T$ and hence $\Sigma$ will not be everywhere spacelike. 

As discussed in the Introduction, it is not a well-posed problem to evolve initial data if $\Sigma$ is not spacelike. Of course, we know that this particular data on $\Sigma$ {\it can} be evolved - the resulting solution is just the black hole solution described above. However, the lack of well-posedness implies that this  procedure is infinitely fine-tuned: if we make a generic (smooth) perturbation to the initial data on $\Sigma$ (for $v$ very close to $1$) then it will not be possible to evolve the perturbed data either forwards or backwards in time. Hence there is a speed limit for small black holes: they cannot be boosted to velocities arbitrarily close to the speed of light.

One might criticise this argument on the grounds that there is no unique way to boost a black hole. One could consider a different surface which is asymptotic to $\Sigma$ but differs in the region $L \ll r \ll \sqrt{\alpha}$ in which $\Sigma$ can fail to be spacelike. However, note that in our argument, the physical metric is actually flat to the level of approximation used because we neglected terms of order $\mu/r^{d-3}$. The boost used above is a symmetry of this flat metric. Therefore our surface $\Sigma$ conforms to the usual idea of a boosted hypersurface in the region relevant for the above argument. 
Of course the effective metrics are not flat to this level of approximation because they include larger terms, of order $\alpha \mu/r^{d-1}=(L/r)^{d-1}$. 

It is interesting to determine the critical value of $v$ below which $\Sigma$ is spacelike. For $\Sigma$ to be spacelike w.r.t. the tensor metric we need $\xi$ to be timelike everywhere, which is true iff\footnote{We used $\theta_1 = \pi/2$ to derive this. It is not hard to show that other values of $\theta_1$ gives less stringent constraints.}
\be
 v^2 < v_{\rm max}^2 \equiv \min \frac{r^2}{\tilde{r}^2 f c_T} \ee
where the minimum is taken over, say, $r > r_H$. For a small black hole, the minimum is achieved for $L \sim r \ll \sqrt{\alpha}$, for which $f \approx 1$ and $\tilde{r} \approx r$ hence
\be
v_{\rm max}^2 \approx \frac{1}{\max c_T} = \frac{1}{3-2\sqrt{1-\frac{1}{(d-4)^{2}}}} < 1.
\ee
This is the speed limit for an arbitrarily small black hole. More generally, $v_{\rm max}$ will depend on the mass of the hole, with $v_{\rm max} \rightarrow 1$ for a large black hole.

It is natural to ask what would happen if one attempted to accelerate a small black hole to a speed greater than $v_{\rm max}$. As emphasized in Ref. \cite{Geroch2010}, one would have to specify the details of how one would attempt to achieve this acceleration using only the fields present in the theory. Perhaps one could set up initial data consisting of several black holes in the hope that a ``gravitational slingshot'' effect could be used to accelerate a small black hole to a speed greater than $v_{\rm max}$. However, as we will see in more detail below, the gravitational interaction associated with small black holes in EGB is very different from GR so there is no reason to believe that a small black hole in such a system would behave in the same was as it would in GR. Whatever the system does, it will not result in a small black hole moving with a speed arbitrarily close to the speed of light at some later time. This is because an ``instant of time'' corresponds to a spacelike hypersurface and the argument above excludes the possibility of a small black hole moving arbitrarily close to the speed of light on such a surface. 

\subsection{Graviton trajectories}

\label{sec:gravitontrajectories}

As discussed above, characteristic hypersurfaces are generated by bicharacteristic curves and, in the present case, these are simply the null geodesics of the effective metrics. Hence, in the geometric optics approximation, the worldlines of high-frequency gravitons are null geodesics of the effective metrics. We will need to determine these geodesics in order to calculate the time delay.
 
\begin{figure}[h!]
	\centering
	\subfigure{\includegraphics[height=5cm]{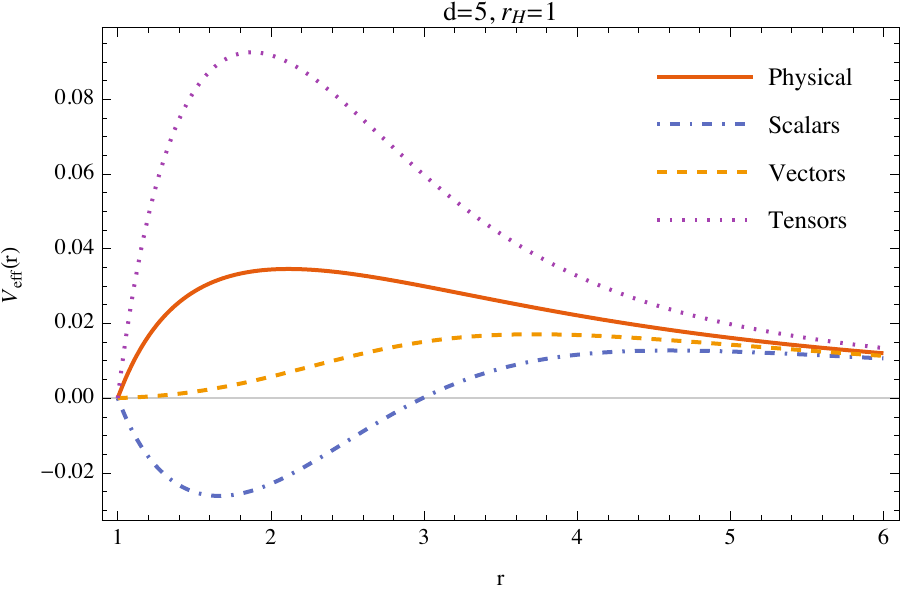}}
	\subfigure{\includegraphics[height=5cm]{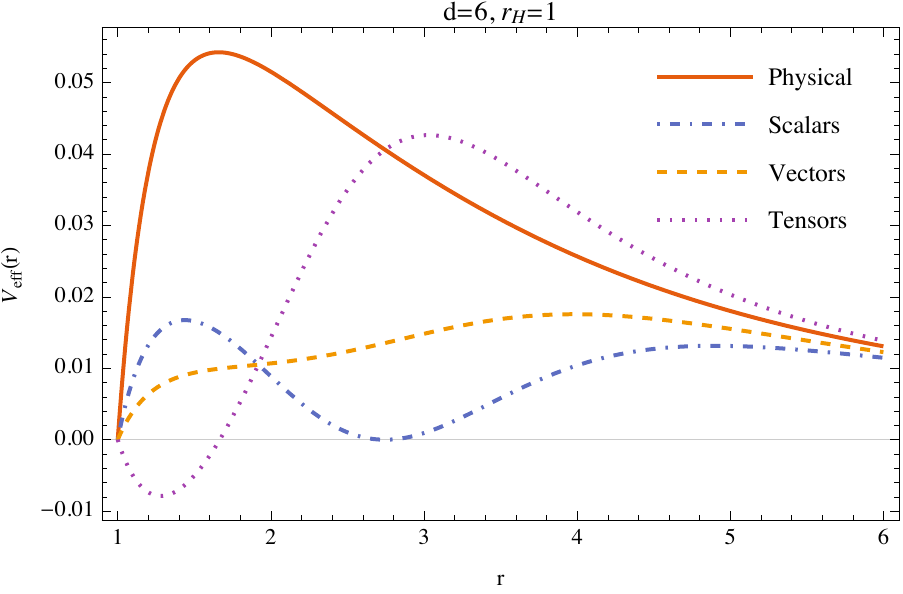}}\\
	\subfigure{\includegraphics[height=5cm]{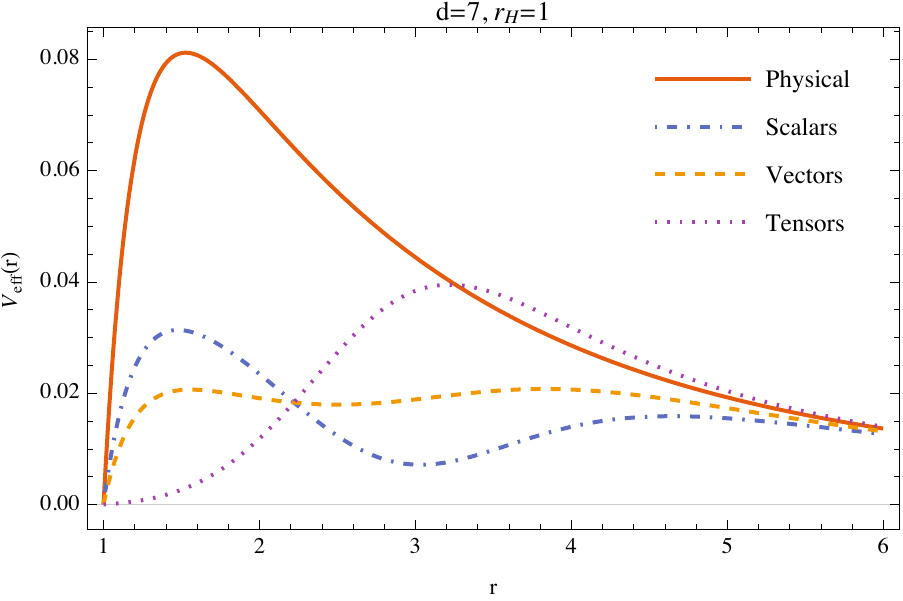}}
	\subfigure{\includegraphics[height=5cm]{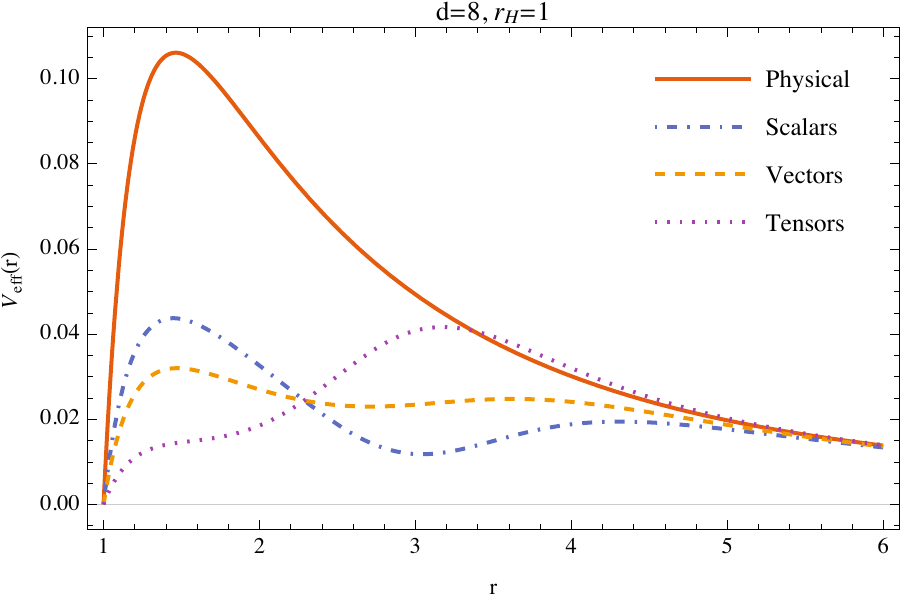}}
	\caption{Effective potentials for a black hole with \(r_{H}=1\) in \(d=5,6,7,8\) dimensions. We fix the Gauss-Bonnet coupling \(\lambda_{\rm GB}=2\). The red curve corresponds to the effective potential for photons, i.e., null geodesics of the physical metric ($c=1$). Superluminal propagation ($c_A>1$) corresponds to an effective potential which is larger than that for photons. This happens only for tensor polarizations. The violation of hyperbolicity is associated with the region in which one of the effective potentials becomes negative. This happens near the horizon for small black holes in five and six dimensions.}
	\label{fig:eff_pot}
\end{figure}

Consider a null geodesic of $G_{Aab}$. Introducing polar coordinates $(\theta_1,\theta_2, \ldots, \theta_{d-3},\phi)$ on $S^{d-2}$, spherical symmetry allows us to assume that the geodesic is confined to the equatorial plane  \(\theta_{1}=\dots=\theta_{d-3}=\pi/2\). Associated to the Killing fields $\partial/\partial t$ and $\partial/\partial \phi$ are the conserved quantities
\begin{align}
	E=f(r)\dot{t}\quad
	J=\frac{r^2}{c_A(r)} \dot{\phi}
\end{align}
where a dot denotes a derivative w.r.t. an affine parameter $\lambda$. $E$ and $J$ are not physical because they depend on the choice of affine parameter. However their ratio is independent of this choice:
\be
 b = \frac{J}{E}
\ee
and this is the impact parameter of the geodesic. The null condition gives
\begin{equation}
	\frac{1}{2}\dot{r}^2+J^2 V^{(A)}_{\rm eff}(r)=\frac{1}{2}E^{2},
\end{equation}
where the effective potential is given by
\begin{equation}
	\label{eq:eff_pot}
	V^{(A)}_{\rm eff}(r)=\frac{f(r)c_{A}(r)}{2r^{2}}.
\end{equation}
Plots of the effective potential for some different cases are given in Ref. \cite{Reall2014} and also in Figure~\ref{fig:eff_pot}.

The effective potentials exhibit a local maximum corresponding to an unstable graviton orbit analogous to the photon sphere in GR. Hence in EGB there is a  distinct ``graviton sphere''  for each graviton polarisation. We will refer to these as the ``scalar sphere'', ``vector sphere'' and ``tensor sphere''. (In some cases, it turns out that there are two local maxima and a local minimum of the effective potential and hence three graviton spheres.) If $V^{(A)}_{\rm max}$ denotes the maximum of the effective potential then a graviton incident from large distance will be absorbed by the black hole if
\be
\label{bmin}
b^2 < \frac{1}{2V^{(A)}_{\rm max}} \qquad \Rightarrow \qquad {\rm absorption}.
\ee
We will consider only gravitons with larger impact parameter, which are scattered by the black hole. 

For $r \gg L$, equations (\ref{fexp}) and (\ref{cexp}) imply that the effective potentials have the expansion
\begin{equation}
\label{effexpansion}
	V_{\rm eff}^{(A)}(r)=\frac{1}{2r^{2}}-\frac{\mu}{2r^{d-1}}+\beta_{A}\frac{\alpha \mu }{r^{d+1}}+\dots
\end{equation}
\noindent The first two terms are familiar from GR: the first is a centrifugal barrier and the second is responsible for the deflection of light rays and the time delay of photons. The third term arises from the Gauss-Bonnet interaction. For the effects of this term to be non-negligible compared to the second term we need $r \lesssim \sqrt{\alpha}$. Since we have assumed $r \gg L$ this requires $L \ll \sqrt{\alpha}$, which implies (\ref{smallbh}), i.e., the black hole has to be small compared to the GB scale for this term to be important.\footnote{This is not possible for $d=5,6$ because of the failure of hyperbolicity for small black holes with $d=5,6$.} Notice that this term is negative for vectors and scalars but positive for tensors. Hence, for a small EGB black hole, tensor-polarized gravitons experience a new {\it repulsive} interaction for $L \ll r \lesssim \sqrt{\alpha}$. It is this repulsive interaction that allows for the possibility of a time advance. 

For a small black hole, the effective potential also simplifies in the region $L \sim r \ll \sqrt{\alpha}$. In this region we can approximate $f \approx 1$ and hence
\be
\label{approxpotential}
 V^{(A)}(r) \approx \frac{c_A(r)}{2r^2}
\ee
which depends only on the length scale $L$. For tensor modes, the RHS typically has a maximum at some value $r \sim L$ so we deduce that the ``tensor sphere'' has $r \sim L$. Since $c_A \sim 1$, we deduce from (\ref{bmin}) that tensor-polarized gravitons which scatter from the black hole (rather than being absorbed) must have $b \gtrsim L$.\footnote{Hence the absorption cross-section for (high frequency) tensor-polarized gravitons by a small black hole scales as $L^{d-2}$ rather than $r_H^{d-2}$.} Note also that the presence of this maximum implies that the interaction between tensor-polarized gravitons and the black hole must become attractive again for $r \lesssim L$.


\section{Time delay and time advance} 
\label{sec:time_delay}

\subsection{Photon time delay in GR}
\label{photons}

As discussed in the Introduction, there is no gauge-invariant definition of the Shapiro time delay applicable to a large class of spacetimes \cite{Gao2000}. However, for a static, spherically symmetric spacetime, there is an unambiguous definition of this quantity \cite{CabreraPalmer2003}. The idea is to compare the time it takes a photon to travel between two points of a spherical cavity with the corresponding time in Minkowski spacetime. In more detail, one can introduce coordinates $(t,r,\theta,\phi)$ (in 4d GR) so that the metric takes the form
\be
 ds^2 = -A(r) dt^2 + B(r) dr^2 + r^2 d\Omega^{2}
\ee
with $A,B>0$. Now one can consider a photon trajectory which starts and ends at $r=R$, with $r \le R$ along the trajectory. Using the spherical symmetry, we can assume that the motion is confined to the equatorial plane $\theta=\pi/2$. Assume it starts at $t=t_0$, $\phi=\phi_0$ and ends at $t=t_0 + \Delta t$, $\phi=\phi_0 + \Delta \phi$. The coordinate time to traverse the cavity is $\Delta t$. The proper time (according to a cavity observer) is
\be
\Delta \tau = \sqrt{A(R)} \Delta t.
\ee
One can compare this with corresponding quantity in Minkowski spacetime where the trajectory is simply a straight line traversing the cavity, which takes proper times
\begin{equation}
\label{Minkdelay}
	\Delta\tau_{\rm Mink}=2 R \sin\left(\frac{\Delta\phi}{2}\right).
\end{equation} 
Hence the time delay can be defined as
\be
D \equiv \Delta \tau - \Delta\tau_{\rm Mink}.
\ee
A simple argument (based on \cite{CabreraPalmer2003}) shows that, in GR, it is impossible to have $D < 0$ for a large class of spacetimes of the above form.
Consider spacetimes for which $A'(r) \ge 0$ and $B(r) \ge 1$. Examples of such spacetimes include the (positive mass) Schwarzschild solution and perfect fluid stars with positive energy density and pressure. Now let $\lambda \in [0,1]$ be an arbitrary parameter along the photon trajectory (which has $r(\lambda) \le R$). Since the photon trajectory is null we have
\be
 \Delta t = \int_0^1 \sqrt{A^{-1} B \dot{r}^2 + A^{-1} r^2 \dot{\phi}^2}\; d\lambda \ge \frac{1}{\sqrt{A(R)}} \int_0^1 \sqrt{ \dot{r}^2 + r^2 \dot{\phi}^2} \; d\lambda \ge \frac{2 R}{\sqrt{A(R)}} \sin\left(\frac{\Delta\phi}{2}\right)
\ee 
where the first inequality follows from $B(r) \ge 1$ and $A'(r) \ge 0$, so $A(r) \le A(R)$ for $r \le R$. The second inequality follows from the fact that the distance in Euclidean space is minimized by a straight line. It follows immediately that $D \ge 0$.\footnote{
If we normalize $t$ so that $A(r) \rightarrow 1$ as $r \rightarrow \infty$ then $A'(r) \ge 0$ implies $A(r) \le 1$ everywhere so $\Delta \tau \le \Delta t$. Hence the time delay defined using $\Delta t$ instead of $\Delta \tau$ also will be positive \cite{CabreraPalmer2003}.}

\subsection{Time delay in EGB} 

\label{sub:time_delay_in_egb}

We can now calculate the Shapiro time delay for gravitons propagating across a spherical cavity in the geometry \eqref{eq:sph_sym_metric}. The cavity is taken to be the surface $r=R$.\footnote{Note that $R$ is the area-radius of the cavity w.r.t. the physical metric, not w.r.t. an effective metric.} Consider a graviton worldline parametrised by $\lambda \in [0,1]$ that has $r \le R$ and starts and ends at $r=R$ with $\phi(0)=0$, $\phi(1)= \Delta \phi$. From the fact that this world line is null w.r.t. the relevant effective metric, the coordinate time $t$ taken for the graviton to traverse the cavity is
\be
\Delta t =  \int_0^1 \sqrt{\frac{\dot{r}^2}{f^2} + \frac{r^2}{f c_A} \dot{\phi}^2}\; d\lambda.
\ee  
One can show that $f$ is monotonically increasing\footnote{
Use $q'=-(d-1)(q^2-1)/(2rq)$ to show that $f'>0$.} so $f(r) \le f(R) \le f(\infty)=1$. Hence
\be
 \Delta t > \frac{1}{\sqrt{f(R)}}  \int_0^1 \sqrt{ \frac{\dot{r}^2}{f} + \frac{r^2}{c_A} \dot{\phi}^2}\; d\lambda > \frac{1}{\sqrt{f(R)}}  \int_0^1 \sqrt{ \dot{r}^2 + \frac{r^2}{c_A} \dot{\phi}^2}\; d\lambda .
\ee
We showed above that $c_S<1$ and $c_V<1$. This implies that for scalar or vector polarizations (or for photons, which have $c_{0}=1$) we have
\be
 \Delta t > \frac{1}{\sqrt{f(R)}}   \int_0^1 \sqrt{ \dot{r}^2 + r^2\dot{\phi}^2}\; d\lambda \ge \frac{2R}{\sqrt{f(R)}} \sin \left( \frac{\Delta \phi}{2} \right) \equiv \frac{\Delta \tau_{\rm Mink}}{\sqrt{f(R)}}
\ee
where, as before, $\Delta \tau_{\rm Mink}$ is the time it takes a photon (or graviton) in Minkowski spacetime to travel across the cavity between the same two points. Converting to proper time we therefore have
\be
 \Delta \tau = \sqrt{f(R)} \Delta t > \Delta \tau_{\rm Mink}
\ee
so the time delay is always positive for scalar or vector polarized gravitons. More physically: gravitons with these polarizations travel slower than photons so, since photons experiences a positive time delay, these gravitons must also experience positive time delay.

The story is different for tensor modes: as we have shown, \(c_{T}(r)\) can be larger than one, so one cannot rule out the possibility of time {\it advance} (e.g. in \(d=5\) we would have \(c_{T}(r)\in[1,3]\) and thus \(\Delta \tau \ge \Delta\tau_{\rm Mink}/\sqrt{3}\)). In the next subsection we will show that time advance is indeed possible, in agreement with \cite{Camanho2014}.

\subsection{Time advance in EGB: perturbative results} 
\label{sub:time_advance_in_egb}

We will now show how one can achieve a negative time delay, i.e., a time advance, for gravitons of tensor polarizations in the space time of a small black hole. We will calculate the time delay explicitly. For completeness, we will also present results for the time delay for vector and scalar graviton polarizations, and also for photons. Consider a graviton trajectory, given by a null geodesic of the relevant effective metric. As before, we assume that this starts at $t=0$, $r=R$, $\phi=0$ and ends at $t=\Delta t$, $r=R$, $\phi=\Delta \phi$ with $r \le R$ along the trajectory. Let \(R_{0}\) be the minimum value of $r$ along the trajectory. As in GR, $R_0$ uniquely labels the trajectory. This is related to the impact parameter $b$ as: 
\begin{equation}
\label{bR0}
	b^2=\frac{R_{0}^{2}}{f(R_{0})c_{A}(R_{0})}.
\end{equation}
We can compute the proper time and the deflection angle from the geodesic equations:
\begin{align}
	\label{eq:dphi}
	\Delta \phi & = 2 b \int_{R_{0}}^{R} dr \; c_{A}(r)\left(r^{2}\sqrt{1-\frac{f(r)c_{A}(r)b^{2}}{r^2}}\right)^{-1},\\
	\label{eq:dtime}
	\Delta t & = 2\int_{R_{0}}^{R} dr  \; \left(f(r)\sqrt{1-\frac{f(r)c_{A}(r)b^{2}}{r^2}}\right)^{-1}.
\end{align}
As before, this includes results for photons (with $c_0=1$). Recall that \(R_{0}\) must be larger than the radius of the \emph{photon/graviton sphere} for the physical/effective metrics. Both the time delay and the deflection angle will diverge as \(R_{0}\) approaches this value since the corresponding trajectories will orbit the black hole many times. 

We will first calculate the above quantities for a graviton trajectory that has $R_0 \gg L$ and also $R_0 \gg \mu^{1/(d-3)}$, and hence $r \gg L$, $r \gg \mu^{1/(d-3)}$ along the trajectory.\footnote{For a large black hole, i.e., $r_H \gg \sqrt{\alpha}$, these conditions reduce to $R_0 \gg r_H$. For a small black hole, i.e., $r_H \ll \sqrt{\alpha}$, they reduce to $R_0 \gg r_{H}^{\frac{d-5}{d-3}}\alpha^{\frac{1}{d-3}}$.} Under these assumptions, $b$ and $R_0$ are related by
\be
 b^2=R_0^2 \left[ 1 + \frac{\mu}{R_0^{d-3}} \left(1-\frac{2\alpha\beta_{A}}{R_0^2} \right) + \ldots \right].
\ee 
We will assume also that the cavity radius is large: $R \gg R_0$. The perturbative calculation of the deflection angle is explained in Appendix \ref{sec:approximation}. The result is
\begin{equation}
	\label{deflectionfiniteR}
	\Delta\phi\approx\pi-2\frac{R_{0}}{R}+\frac{\mu}{R_0^{d-3}} \sqrt{\pi} (d-1)\left(1-\frac{2\alpha\beta_{A}}{R_0^2}\frac{(d-2)}{(d-1)}\right)\frac{\Gamma\left(\frac{d}{2}\right)}{2\Gamma\left(\frac{d+1}{2}\right)}.
\end{equation}
This has a well-defined limit when the cavity radius is taken to infinity at fixed $R_0$. Using $b \approx R_0$ to write the result in terms of $b$ gives
\begin{equation}
	\label{eq:dphi_approx}
	\Delta \phi_\infty \equiv \lim_{R\rightarrow \infty} \Delta \phi \approx\pi+\frac{\mu}{b^{d-3}} \sqrt{\pi} (d-1)\left(1-\frac{2\alpha\beta_{A}}{b^2}\frac{(d-2)}{(d-1)}\right)\frac{\Gamma\left(\frac{d}{2}\right)}{2\Gamma\left(\frac{d+1}{2}\right)}.
\end{equation}
This is the analogous to the result for the deflection of light by the Schwarzschild solution in GR. Note that $\Delta \phi_\infty>\pi$ if $\beta_A \le 0$ so scalar and vector polarised gravitons, and photons, are always deflected towards the black hole. However, for tensor polarised gravitons, since $\beta_T>0$, we see that $\Delta \phi_\infty<\pi$ when
\be
\label{repulsion}
 b <  \sqrt{\frac{2(d-2)\alpha}{(d-3)(d-4)}}.
\ee
This is consistent with our previous assumptions if, and only if, (\ref{smallbh}) holds, i.e., iff the black hole is small. Hence, for a small black hole, tensor-polarized gravitons with $b$ obeying (\ref{repulsion}) (and $b\approx R_0 \gg L$ from our previous assumptions) are deflected {\it away} from the black hole. This is precisely because such gravitons experience the repulsive short-distance interaction in (\ref{effexpansion}) that we discussed above. 

Using the same approximations as above we find that the time it takes a graviton to cross the cavity is
\be
 \Delta t = 2R + \frac{\mu}{R_{0}^{d-4}}\left[\sqrt{\pi}\left(1-\frac{2\alpha\beta_{A}}{R_{0}^{2}}\frac{(d-4)}{(d-3)}\right)\frac{(d-1)(d-3)}{(d-4)}\frac{\Gamma\left(\frac{d}{2}\right)}{2\Gamma\left(\frac{d+1}{2}\right)}\right] +\ldots
\ee
and $R \gg R_0$ implies that $\Delta \tau = \Delta t$ to this level of approximation. The corresponding time in flat spacetime, with deflection angle $\Delta \phi$ is given by \eqref{Minkdelay}, which can be written as
\be
 \Delta \tau_{\rm Mink} = 2 R \cos \left( \frac{\pi - \Delta \phi}{2} \right) = 2R \left[ 1 - \frac{1}{2}  \left( \frac{\pi - \Delta \phi}{2} \right)^2 + \ldots \right]
\ee
so plugging in our perturbative result \eqref{deflectionfiniteR} gives
\begin{align}
 \Delta \tau_{\rm Mink} = & \: 2R + \frac{\mu}{R_{0}^{d-4}}\left[\sqrt{\pi}\left(1-\frac{2\alpha\beta_{A}}{R_{0}^{2}}\frac{(d-2)}{(d-1)}\right)(d-1)\frac{\Gamma\left(\frac{d}{2}\right)}{2\Gamma\left(\frac{d+1}{2}\right)}\right]\\
 & + {\cal O}\left( R\left( \frac{ \mu}{R_0^{d-3}} + \frac{ L^{d-1}}{R_0^{d-1}}\right)^2 \right).
 \end{align}
Note that second order corrections to this result grow linearly with $R$. For these to be small compared to the terms that we have retained we need 
\be
\label{nottoobig}
 \frac{R}{R_0} \ll \frac{R_0^{d-3}}{\mu}, \qquad \frac{R}{R_0} \ll \left( \frac{R_0}{L}\right)^{d-1}=\frac{R_0^{d-1}}{\alpha \mu}
\ee
i.e. the cavity is large ($R/R_0 \gg 1$) but not too large. If these assumptions are not satisfied then most of the trajectory is in a region where spacetime is almost flat and a large positive time delay (proportional to $R$) results simply because, in this flat region, there is a shorter (straight line) path available which remains far from the black hole.

Combining the above results gives the time delay as (using $R_0 \approx b$)
\be
\label{advance}
 D=\Delta \tau - \Delta \tau_{\rm Mink} \approx\frac{\mu}{b^{d-4}}\left[\sqrt{\pi} \left(1-\frac{2\alpha\beta_{A}}{b^2}\frac{(d-4)}{(d-1)}\right)\frac{(d-1)}{(d-4)}\frac{\Gamma\left(\frac{d}{2}\right)}{2\Gamma\left(\frac{d+1}{2}\right)}\right].
\ee
We see that scalar or vector polarised gravitons, or photons, suffer a positive time delay as expected. However, for tensor polarised gravitons, a negative time delay, i.e. a time {\it advance} results when
\begin{equation}
	b<\sqrt{\frac{2 \alpha}{(d-3)}}.
\end{equation}
Since $b \approx R_0$, this is consistent with $R_0 \gg L$, as assumed above, only for a small black hole. For such a black hole, stated in terms of $b$, our assumptions in deriving (\ref{advance}) are
\be
 b \gg L, \qquad 1 \ll \frac{R}{b} \ll \frac{b^{d-3}}{\mu}, \left( \frac{b}{L} \right)^{d-1}.
\ee
Note that this overlaps the region for which $\Delta \phi_\infty < \pi$, i.e., the gravitons that experience a time advance are also deflected away from the black hole. Both effects arise from the repulsive term in the effective potential discussed above. 

As discussed above, we need to impose an upper bound on $R/b$ to see the time advance since for very large $R$ there will be a large time delay, proportional to $R$, which occurs because the trajectory has undergone a deflection. However, for the special case of a trajectory which saturates (\ref{repulsion}), we have $\Delta \phi_\infty = \pi$, i.e., there is no net deflection (the effect of the short-distance repulsion is cancelled by the effect of the long-distance attraction). In this case, we no longer need to impose an upper bound on $R$: it is easy to see that the above derivation holds for arbitrarily large $R$. Hence the result (\ref{advance}) is valid for $R\rightarrow \infty$ in this special case. It is easy to see that, for this value of $b$, the expression (\ref{advance}) is positive, so this special trajectory experiences a time delay. Hence, in this special case, we have a gauge-invariant definition of the time delay for a graviton propagating in from infinity and returning to infinity.


\subsection{Time advance in EGB: numerical results} 
\label{sub:numerics}

The above perturbative calculation demonstrates that a time advance is possible for tensor-polarized gravitons propagating in the geometry of a small black hole for $d \ge 7$. However, several questions remain. As discussed above, small black holes are unphysical for $d=5,6$. So for $d=5,6$ we will have to study black holes which are not small in order to demonstrate that a time advance is possible. Furthermore, we would like to determine (for any $d$) how large the time advance can be. The perturbative result indicates that the time advance increases as the impact parameter decreases so we would like to consider $b$ as small as possible. The lower bound on the impact parameter is $b \sim L$ but the above calculation assumes $b \gg L$. Hence to determine the largest possible time advance we will need to use a different method. Finally, we discussed above the case of special trajectories which experience no net deflection, for which the time delay is finite as $R \rightarrow \infty$. We saw that, the resulting time delay is always positive when the perturbative calculation is valid. But what about trajectories with $b \sim L$? Could these exhibit zero net deflection? If so, can they exhibit a time advance?

To address the above questions, we will resort to numerical integration. We will compute numerically both the deflection angle \eqref{eq:dphi} and the time delay \eqref{eq:dtime} for the tensor modes as functions of the impact parameter and plot the results for different parameters. In practice, we do this calculation by using $R_0$, the minimum value of $r$, to label the trajectory and determine $b$ from $R_0$ using (\ref{bR0}). 

\begin{figure}
	\centering
	\subfigure{\includegraphics[height=5cm]{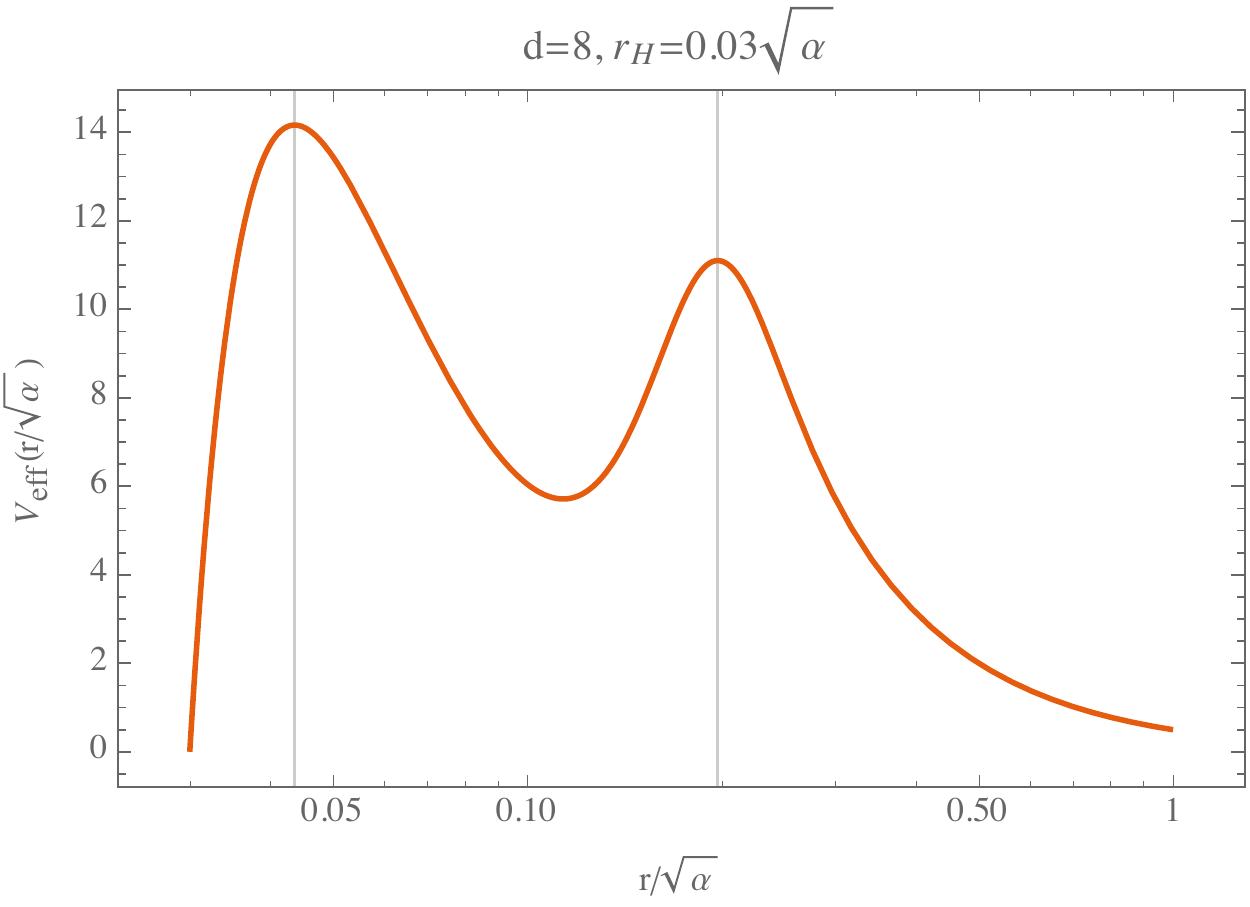}}
	\subfigure{\includegraphics[height=5cm]{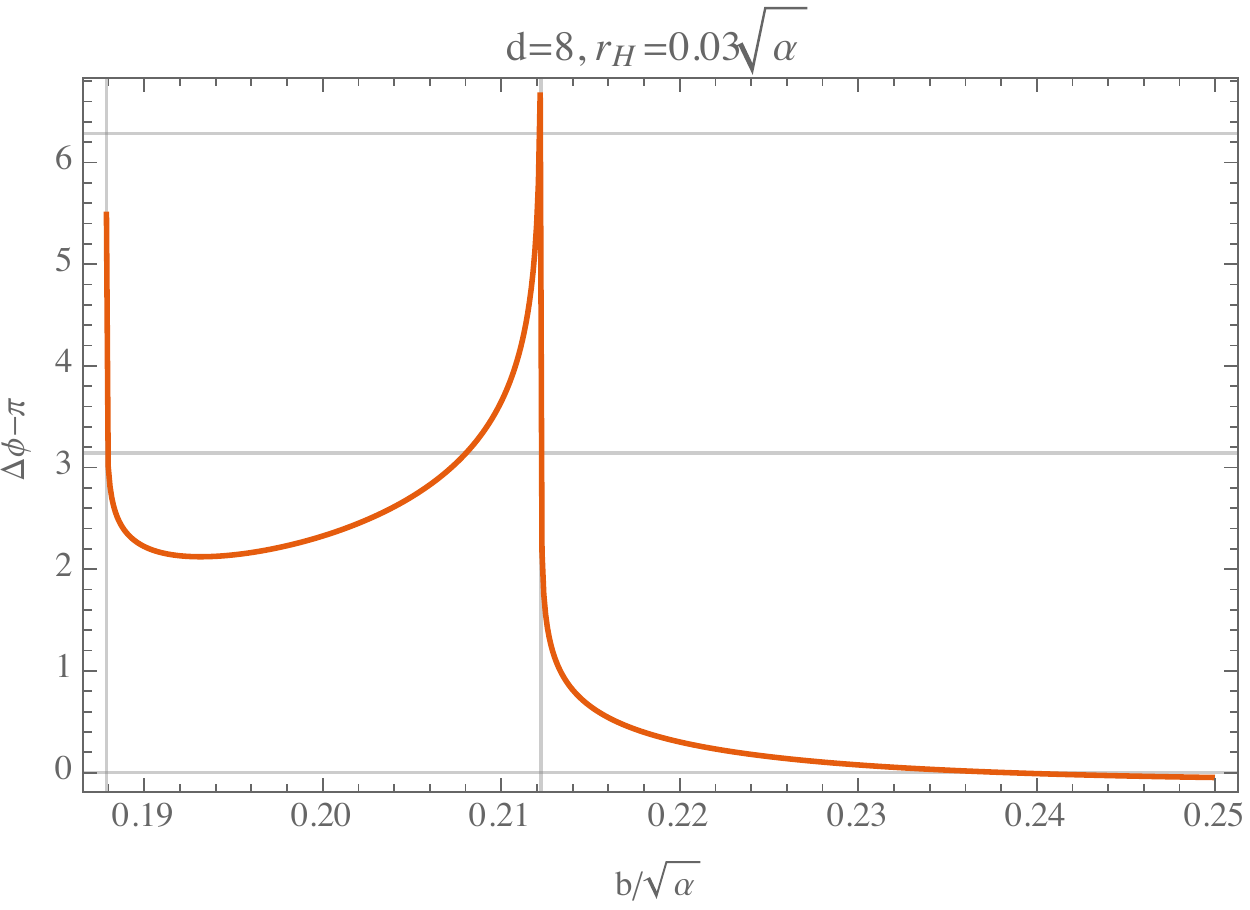}}
	\caption{Effective potential (\emph{left}) and deflection angle (\emph{right}) for tensor polarized gravitons scattered by a small black hole in \(d=8\). We set \(r_{H}=0.03 \sqrt{\alpha}\), which gives \(\mu\approx1.4\times10^{-5}\alpha^{5/2}\) and \(L\approx0.2\sqrt{\alpha}\).}
	\label{fig:8d_angle}
\end{figure}

We start by calculating the deflection angle for tensor-polarized gravitons by a small black hole with $d \ge 7$, in the limit of infinite cavity radius. We will compute the deflection angle \(\Delta\phi_{\infty}\) as a function of the impact parameter \(b\). For $d \ge 8$, complicated behaviour arises at small $b$ because the tensor effective potential for small black holes has a complicated form with a local minimum and two local maxima - see Figure~\ref{fig:8d_angle}. The local minimum corresponds to {\it stable} circular graviton orbits around the black hole. The maxima correspond to unstable circular graviton orbits. Hence there are three ``tensor spheres''. Only the local maxima are relevant for scattering of (high frequency) gravitons. As noted above, the deflection angle must diverge at impact parameters corresponding to these maxima and this can be seen in in Fig.~\ref{fig:8d_angle}. This figure also shows that $\Delta \phi_\infty>\pi$ when $R_0$ lies between the two maxima.

\begin{figure}
	\centering
	\subfigure{\includegraphics[height=5cm]{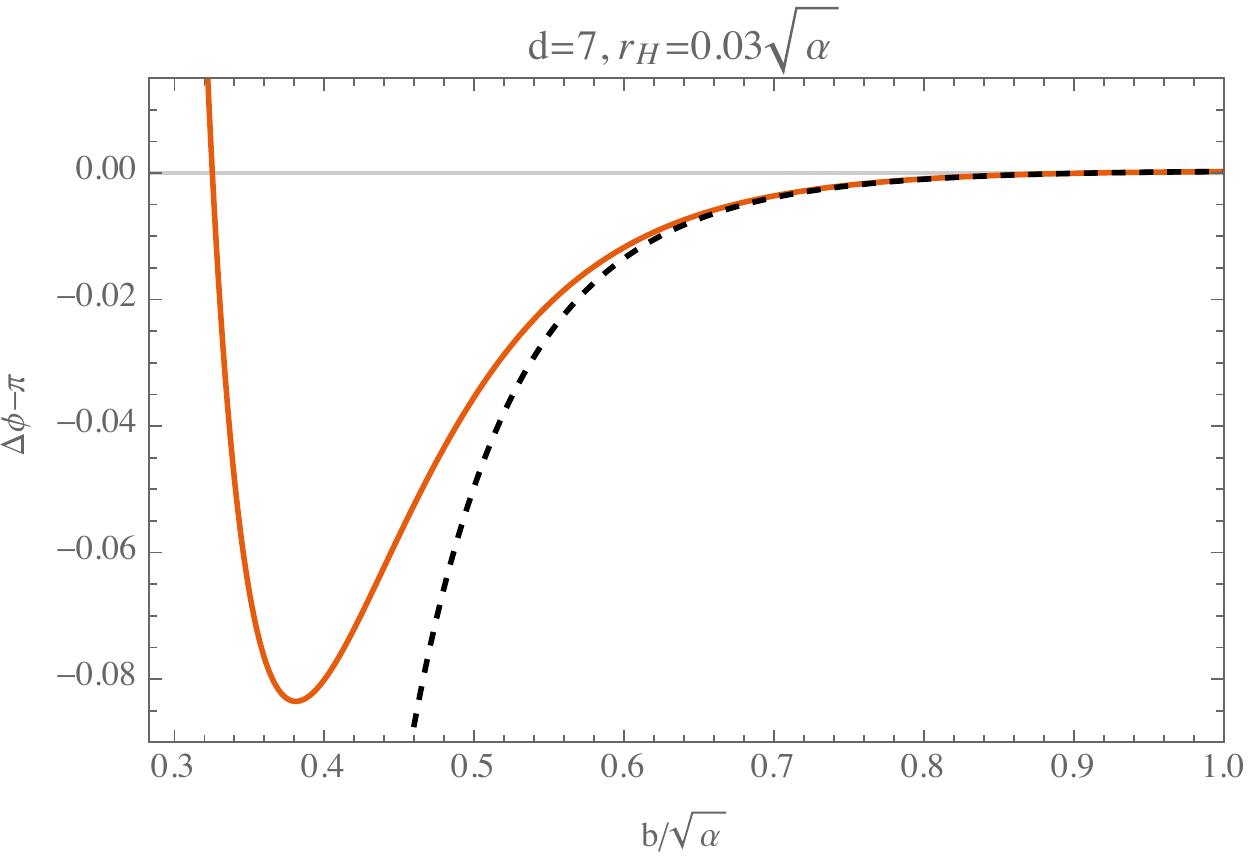}}
	\subfigure{\includegraphics[height=5cm]{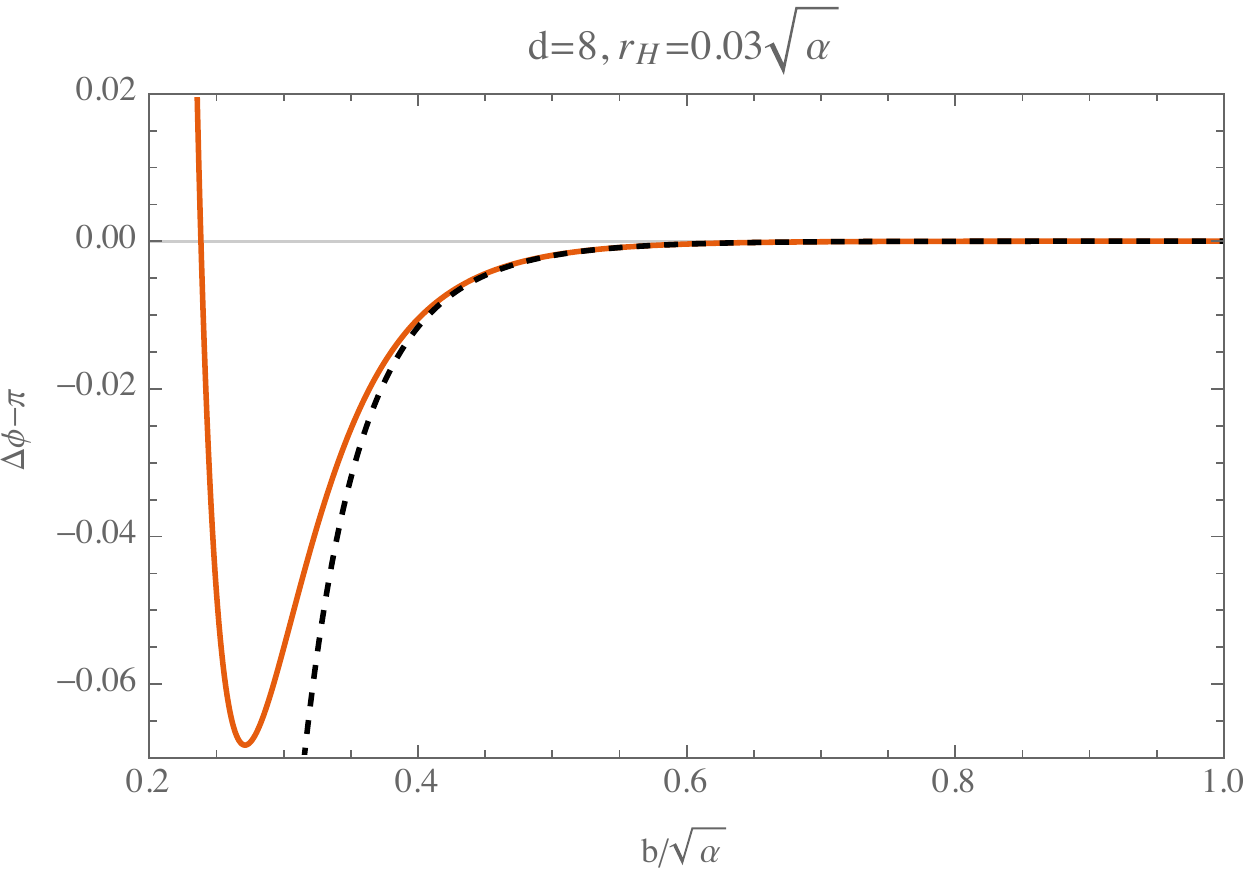}}
	 \caption{Deflection angle for small black holes in \(d=7,8\). We set \(r_{H}=0.03 \sqrt{\alpha}\) which gives \(\mu\approx4.5\times10^{-4}\alpha^{2}\), \(L\approx0.28\sqrt{\alpha}\) in \(d=7\), and \(\mu\approx1.4\times10^{-5}\alpha^{5/2}\), \(L\approx0.2\sqrt{\alpha}\) in \(d=8\). The dashed line represents the perturbative approximation \eqref{eq:dphi_approx}.}
	\label{fig:angle}
\end{figure}

Figure~\ref{fig:angle} shows plots of $\Delta\phi_\infty$ for $d=7,8$ outside the outer tensor sphere. At large $b$, our perturbative results show that $\Delta \phi_\infty>\pi$ although this is not apparent from the plots because $\Delta \phi_\infty -\pi$ is very small. As $b$ decreases, our perturbative result shows that $\Delta\phi_\infty - \pi$ becomes negative, as seen in the plots. The plots show that $\Delta \phi_\infty - \pi$ decreases to a negative minimum and then increases, becoming positive as $b$ is decreased further. This lies outside the validity of the perturbative calculation. Note that there are two values of $b$ for which $\Delta \phi_\infty  = \pi$, i.e., for which there is no net deflection. The larger of these, with $b \gg L$ is encompassed by our perturbative approximation, and we showed above that it gives a trajectory with a positive time delay in the infinite cavity limit. However, we will show that the smaller value of $b \sim L$ can give a time advance in the infinite cavity limit.  

\begin{figure}
	\centering
	\subfigure{\includegraphics[height=5cm]{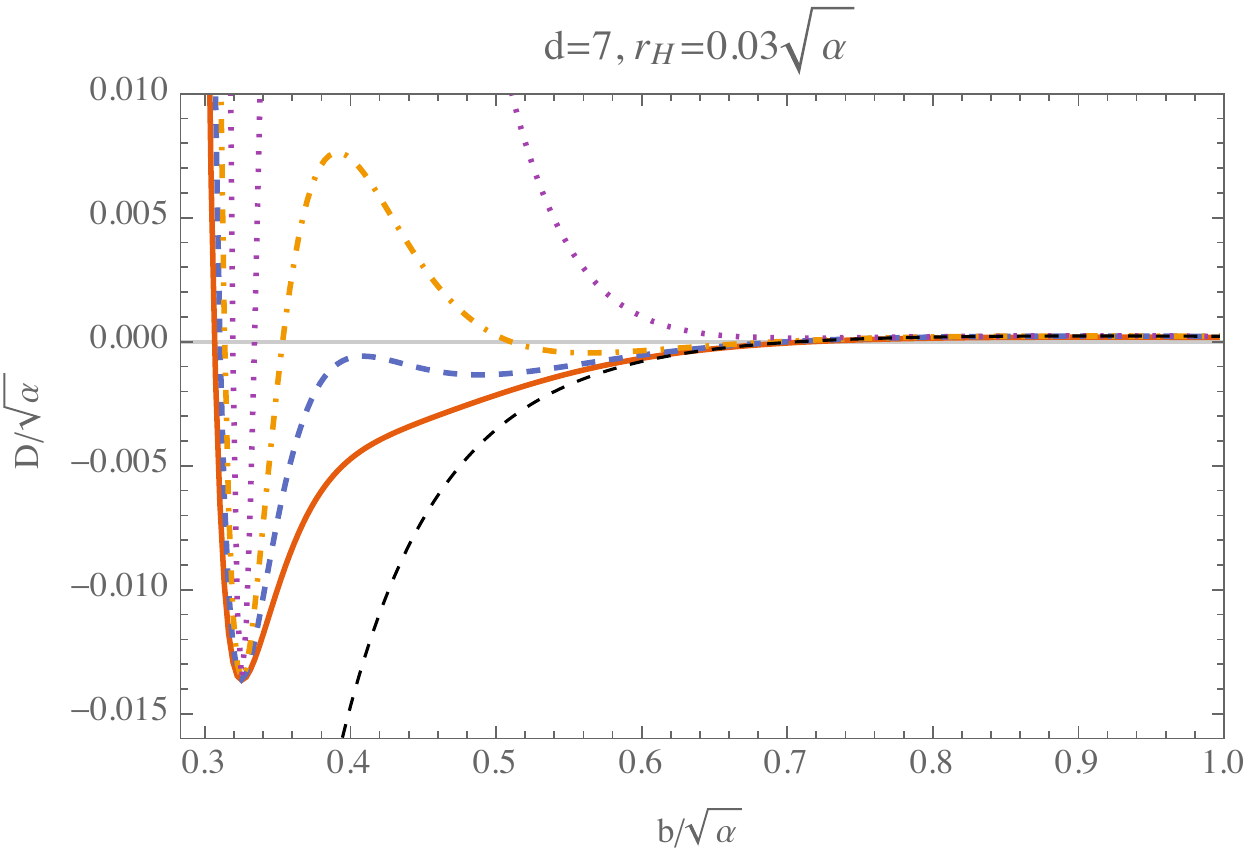}}
	\subfigure{\includegraphics[height=5cm]{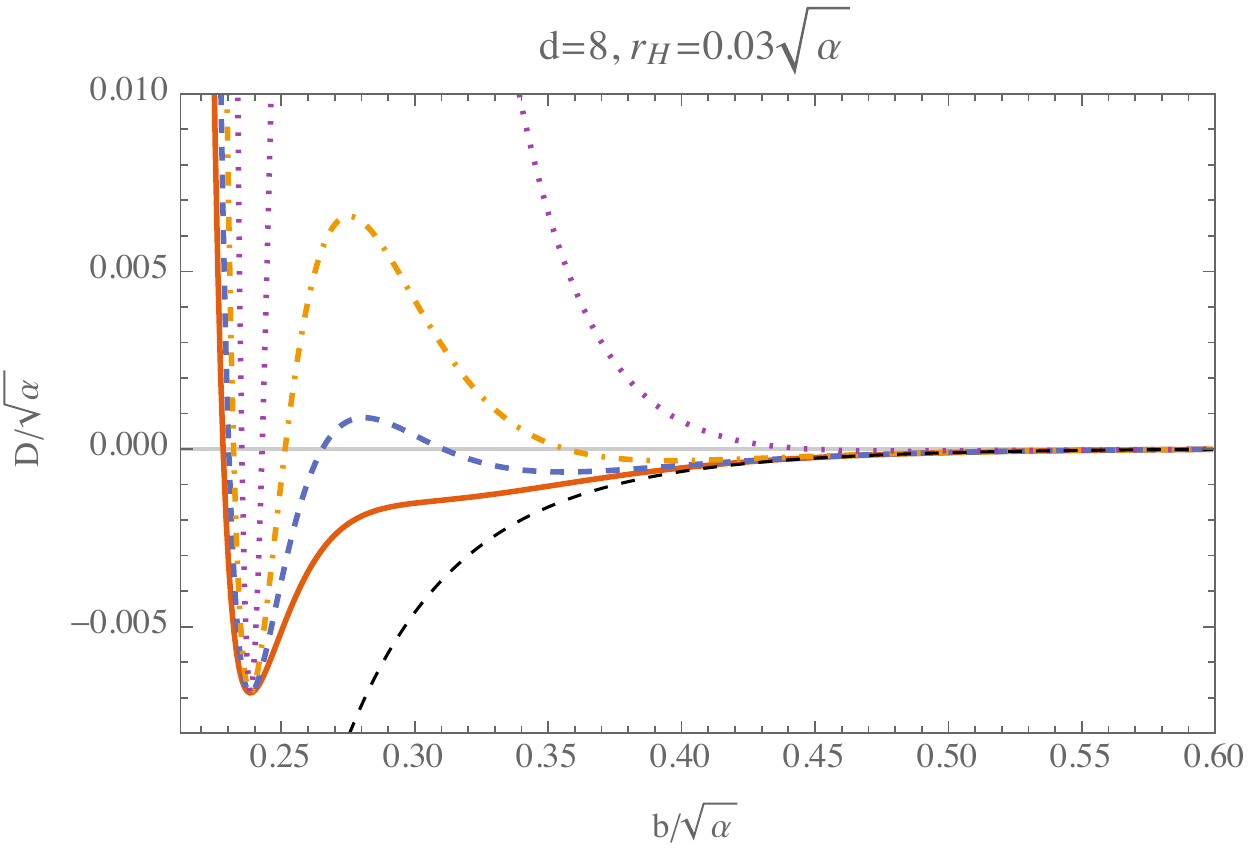}}
	\caption{Time delay for small black holes in \(d=7,8\). We set \(r_{H}=0.03 \sqrt{\alpha}\) as before. The solid, dashed, dot-dashed and dotted lines correspond to \(R=2.5\sqrt{\alpha}\), \(R=5\sqrt{\alpha}\), \(R=10\sqrt{\alpha}\) and \(R=50\sqrt{\alpha}\) respectively. The black dashed line corresponds to the perturbative approximation \eqref{advance}.}
	\label{fig:td}
\end{figure}

Figure~\ref{fig:td} plots the time delay for different values of the cavity radius $R$ with $d=7,8$.\footnote{For $d=8$ we only show results for $R_0$ outside the outer tensor sphere. The time delay is positive for $R_0$ between the two maxima of the effective potential.} At large $b$, the time delay is small and positive, but becomes negative as $b$ is decreased, i.e. there is a time advance as predicted by our perturbative calculation. The size of the time advance increases as $b$ is decreased further but at small enough $b$, the time delay becomes positive again. As expected, increasing $R$ tends to increase the time delay. The only trajectories for which this does not happen correspond to the two special values of $b$ for which the trajectory does not undergo a net deflection. Hence as $R$ is increased, the minimum in these plots, corresponding to the largest time advance, becomes more and more localized around the smaller value of $b$ for which $\Delta \phi_\infty = \pi$. 

 \begin{figure}
	\centering
	\includegraphics[height=5cm]{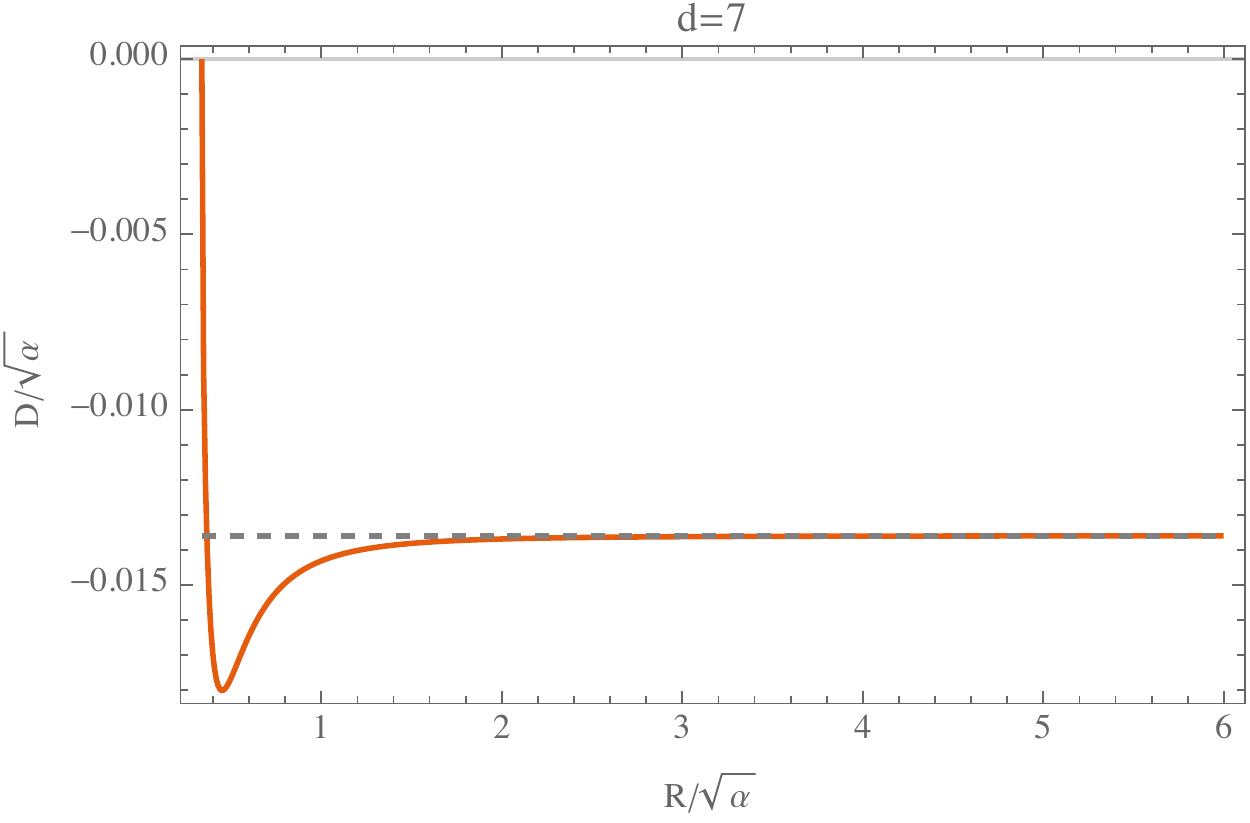}
	\caption{Time delay for fixed \(b\) (such that \(\Delta\phi_{\infty}=\pi\)) expressed as a function of the radius of the cavity for a small black hole in \(d=7\). We set \(r_{H}=0.03 \sqrt{\alpha}\) as before. The dashed line corresponds to the limit \(R \rightarrow \infty\).}
	\label{fig:td_vs_R}
\end{figure}

\begin{figure}
	\centering
	\subfigure{\includegraphics[height=5cm]{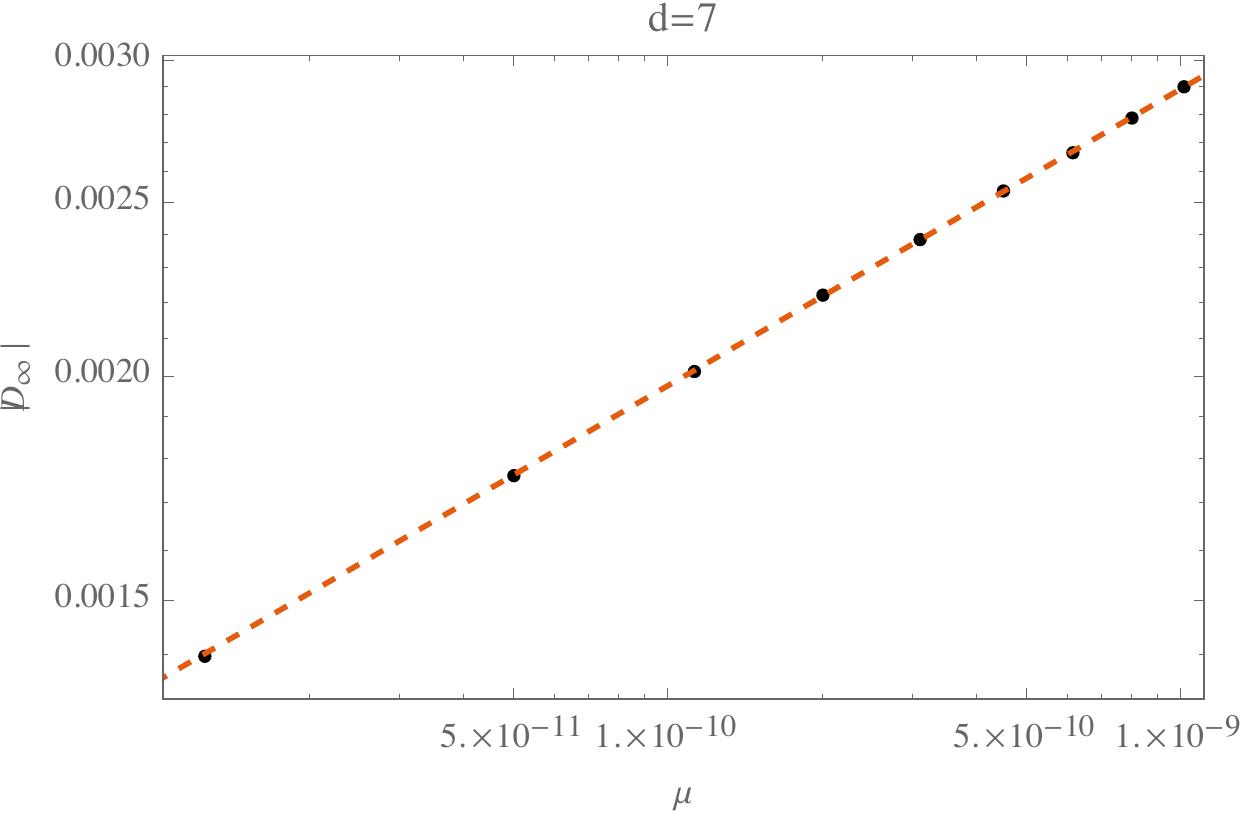}}
	\subfigure{\includegraphics[height=5cm]{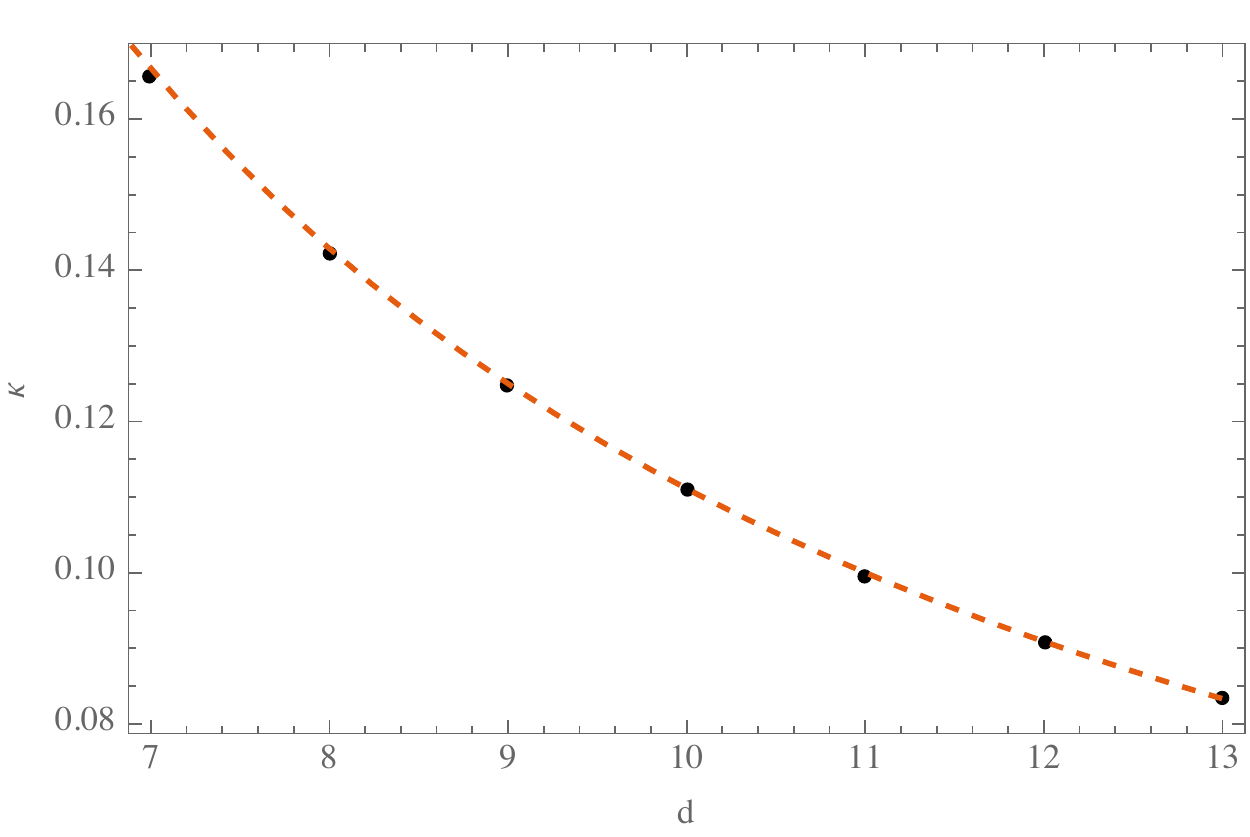}}
	\caption{(\emph{Left}) Log-log plot of the absolute value of the time advance for an undeflected geodesic in \(d=7\) in the limit \(R \rightarrow \infty\) against the mass parameter \(\mu\) (units $\alpha=1$). The dashed line is given by \(a \mu^\kappa\) with \(a\approx0.09\)  and \(\kappa\approx0.165\). 
	(\emph{Right}) Plot of \(\kappa\) against \(d\). The dashed line is \(\frac{1}{d-1}\).} 
	\label{fig:exponent}
\end{figure}

Figure~\ref{fig:td_vs_R} shows how the maximum time advance, corresponding to the minimum in Fig.~\ref{fig:td}, behaves as $R$ is increased. As $R \rightarrow \infty$ we see that the (negative) time delay converges to a finite limit $D_\infty$, as expected for a trajectory with zero net deflection. We would like to understand what scale determines the amplitude of \(D_{\infty}\). The obvious guess is the scale $L$ and this turns out to be correct. In Figure~\ref{fig:exponent} we plot \(\abs{D_{\infty}}\) against the mass parameter \(\mu\) for small black holes in \(d=7\). From the plot we deduce that the relation should be a power law: \(\abs{D_{\infty}}\sim \mu^{\kappa}\) (in units $\alpha=1$). By estimating the value of \(\kappa\) in different dimensions (Figure~\ref{fig:exponent}) we obtain numerically \(\kappa\approx \frac{1}{d-1}\). Recalling that \(L\sim\mu^{1/(d-1)}\) (since $\alpha=1$), we have found:
\begin{equation}
	\abs{D_{\infty}}\sim L .
\end{equation}

One can explain this result analytically as follows. Denote by \(b_{*}\) the value of \(b\) close to the tensor sphere (i.e. \(b_{*}\sim L\)) for which \(\Delta\phi_{\infty}=\pi\). From Fig.~\ref{fig:td_vs_R} it is clear that the most of the time advance arises from the region $r \sim L$. For a small black hole, we have $f \approx 1$ in this region (see end of section \ref{sec:gravitontrajectories}). 
This suggests that we can calculate $D_{\infty}$ for a small black hole by approximating $f=1$ in the integral for the time delay. If we do this then the integrand now varies only over the scale $L$ and so all quantities in the problem are of order $L$. Hence, by dimensional analysis, $D_{\infty}$ must be proportional to $L$. To get an idea of the error made in setting $f=1$ we note that this approximation eliminates the usual GR time delay effect. We can estimate the error made by our approximation by estimating the size of this delay as $\mu/b_*^{d-4} \sim \mu/L^{d-4} =L^3/\alpha$. For a small black hole $L \ll \sqrt{\alpha}$ hence the error is parametrically smaller than the scale $L$ and therefore negligible.

In summary, we have shown that, for a small black hole with $d \ge 7$,  there is a tensor-polarized graviton trajectory with impact parameter $b \sim L$ that experiences no net deflection and, in the infinite cavity limit, experiences a finite time advance of order $L$.\\

\begin{figure}
	\centering
	\subfigure{\includegraphics[height=5cm]{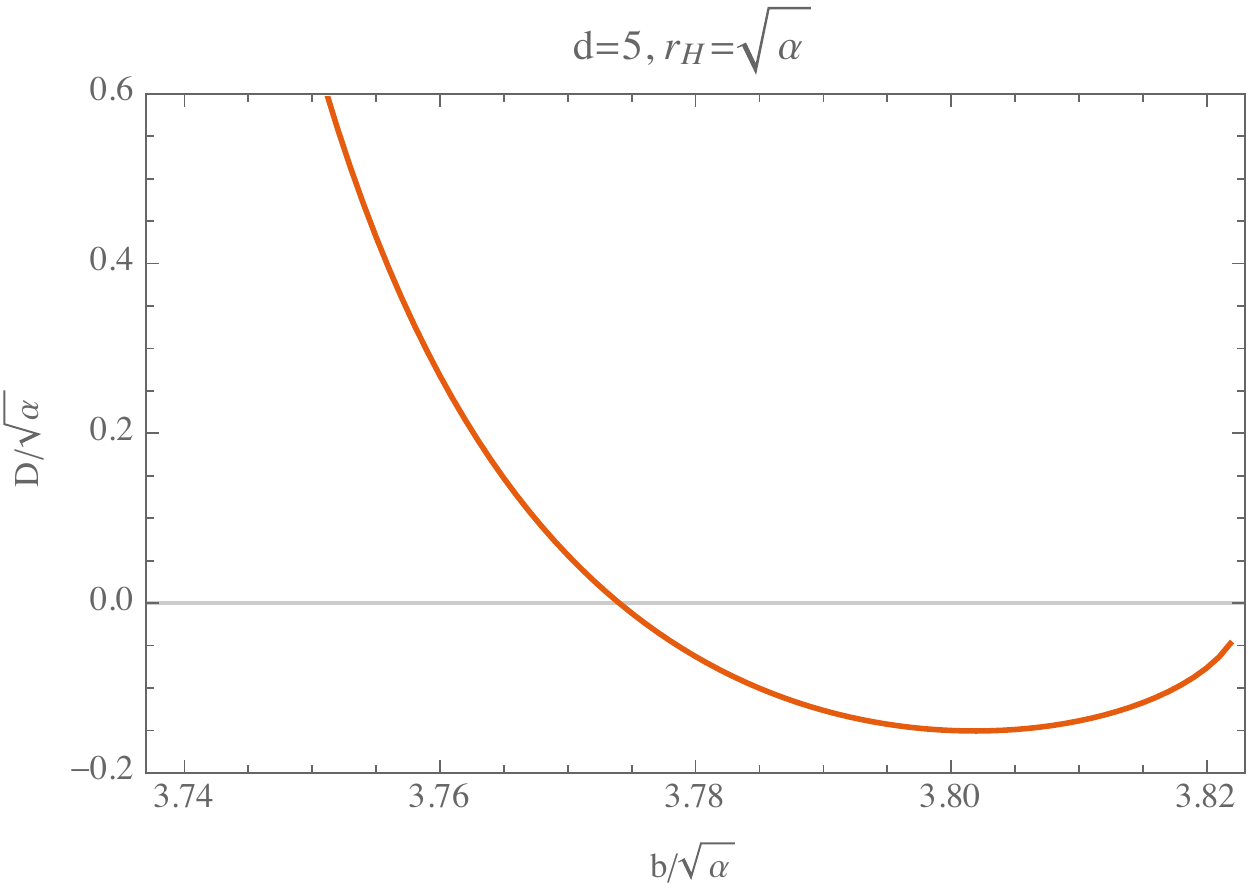}}
	\subfigure{\includegraphics[height=5cm]{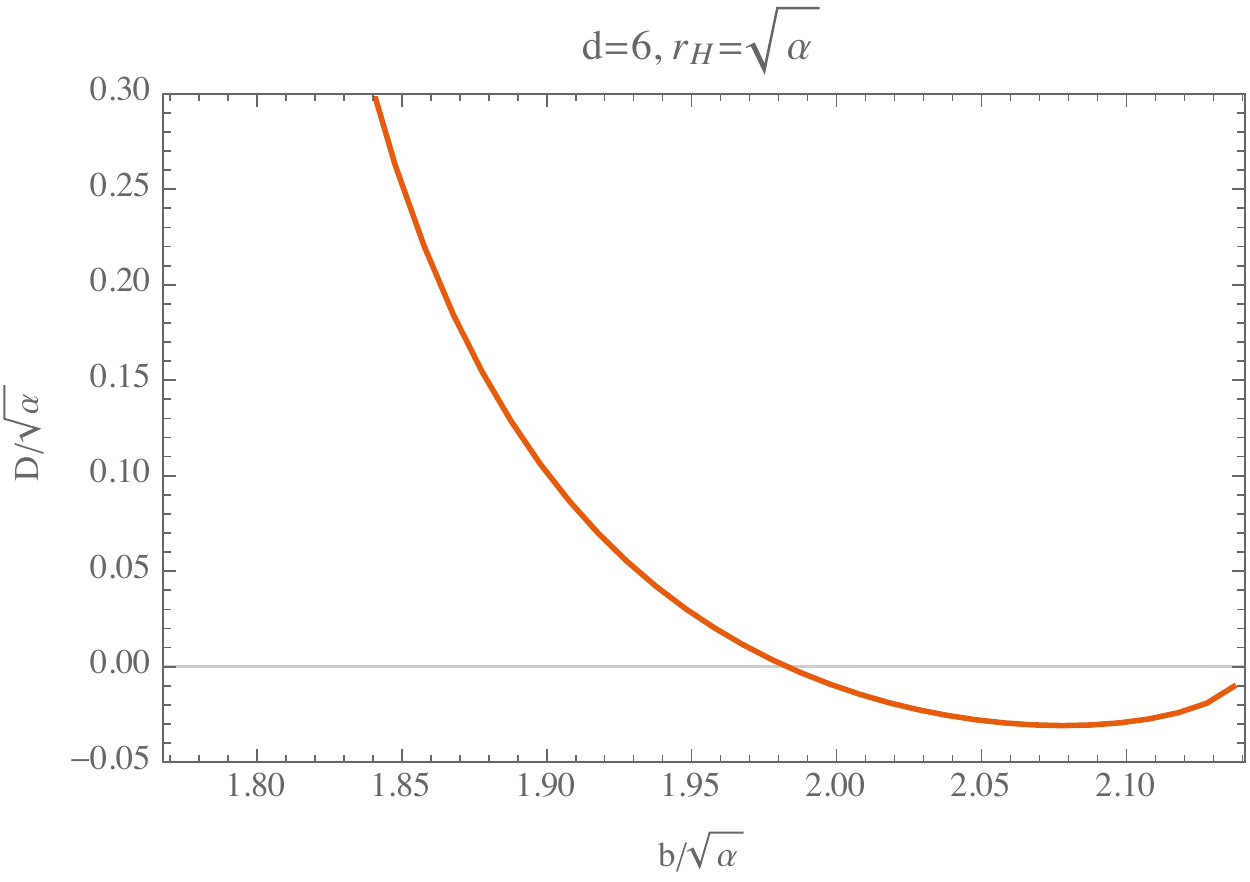}}
	\caption{Time delay for tensor-polarized gravitons in \(d=5,6\) dimensions. In \(d=5\) we set \(r_{H}=2 \sqrt{\alpha}\), which gives \(\mu=4.5\alpha\), \(L\approx1.5\sqrt{\alpha}\), and we choose \(R=3\sqrt{\alpha}\). In \(d=6\) we set \(r_{H}=\sqrt{\alpha}\), which gives \(\mu=1.5\alpha^{3/2}\), \(L\approx1.1\sqrt{\alpha}\), and we choose \(R=2\sqrt{\alpha}\).}
	\label{fig:td_5_6}
\end{figure}

Finally we consider the cases of $d=5,6$ for which we cannot consider arbitrarily small black holes because of the failure of hyperbolicity.\footnote{In \(d=5\) the theory is hyperbolic in the exterior of the black hole for \(r_{H}/\sqrt{\alpha}>\sqrt{1+\sqrt{2}}\approx 1.6\), while in \(d=6\) this happens for \(r_{H}/\sqrt{\alpha}>\left(\sqrt{5 \left(5+2 \sqrt{6}\right)}-1\right)^{-1/2}\approx 0.4\)} We want to show that a negative time delay is possible for $d=5,6$. To do this, consider the case in which \(R=R_{0}(1+\epsilon)\) with $\epsilon \ll 1$. Under the change of variable \(r=R_{0}(1+x)\) the integral for \(\Delta t\) becomes:
\begin{equation}
	\Delta t = 2 R_{0}\int_{0}^{\epsilon}\dd x\; \left(f(R_{0}(1+x))\sqrt{1-\frac{f(R_{0}(1+x))c_{A}(R_{0}(1+x))b^{2}}{R_{0}^2(1+x)^{2}}}\right)^{-1}.
\end{equation}
\noindent We can now expand in powers of \(x\ll1\) and integrate:
\begin{equation}
	\Delta t= \frac{4 R_{0}\sqrt{\epsilon}}{\sqrt{f(R_{0})}\sqrt{2-\frac{R_{0}f'(R_{0})}{f(R_{0})}-\frac{R_{0}c_{A}'(R_{0})}{c_{A}(R_{0})} }}+{\cal O}(\epsilon^{3/2}).
\end{equation}
\noindent Similarly one can compute the deflection angle:
\begin{equation}
	\Delta \phi = \frac{4 \sqrt{c_{A}(R_{0})}\sqrt{\epsilon}}{\sqrt{f(R_{0})}\sqrt{2-\frac{R_{0}f'(R_{0})}{f(R_{0})}-\frac{R_{0}c_{A}'(R_{0})}{c_{A}(R_{0})} }}+{\cal O}(\epsilon^{3/2}).
\end{equation}
\noindent It follows that the time delay is given by:
\begin{equation}
	D= \frac{4 R_{0}\sqrt{\epsilon}}{\sqrt{f(R_{0})}\sqrt{2-\frac{R_{0}f'(R_{0})}{f(R_{0})}-\frac{R_{0}c_{A}'(R_{0})}{c_{A}(R_{0})} }}\left(1-\sqrt{c_{A}(R_{0})}\right)+{\cal O}(\epsilon^{3/2}).
\end{equation}
Since \(c_{S},c_{V}<1\) we see that in this setting the time delay is always positive for scalar and vector modes, as expected. For tensor modes this can be negative. In particular, in \(d=5\) we have \(c_{T}>1\) everywhere and thus, for a black hole of arbitrary size we have a negative time delay when \(R=R_{0}(1+\epsilon)\). In \(d=6\), for \(r\gtrsim L\) we also have \(c_{T}>1\). Motivated by this, we compute numerically the time delay in \(d=5,6\) for values of \(R\) comparable to \(R_{0}\). The numerics confirm that it is possible to obtain a time advance when \(d=5,6\) (Figure~\ref{fig:td_5_6}). We have also studied the deflection angle, which we find is always greater than $\pi$, so zero net deflection trajectories do not occur for $d=5,6$.

\section{Time machines} 
\label{sec:time_machines}
In this section we will discuss the suggestion that one can exploit the negative Shapiro time delay to construct a causality violating spacetime, i.e., a ``time machine'', in Einstein-Gauss-Bonnet theories \cite{Camanho2014}. This argument is closely related to arguments applying to any Lorentz covariant field theory with superluminal propagation, some of which have appeared in Ref. \cite{Adams2006}, which considers various flat space field theories with superluminal propagation. Such time machine constructions have been criticized by Geroch \cite{Geroch2010} (see also Ref.~\cite{Bruneton2007}). In this section we will discuss how these criticisms apply to the constructions of Refs \cite{Camanho2014,Adams2006}. 

Consider first the case of General Relativity. There are many solutions of the Einstein equation which exhibit causality violation e.g. Minkowski spacetime with a periodically identified time direction. We do not reject GR as a physical theory because it admits such solutions. This is because one cannot ``make'' these time machines starting from initial data. Stated mathematically: such causality-violating solutions are not the Cauchy development of any initial data.\footnote{More generally, the class of spacetimes of interest in GR is the class of spacetimes that arises as the maximal Cauchy development of suitable initial data, where ``suitable'' depends on the physical situation e.g. one would usually require the initial data to be geodesically complete and impose some asymptotic boundary condition e.g. asymptotic flatness. By definition, the maximal development is globally hyperbolic and so it can never violate causality. But the maximal development might be {\it extendible} beyond a Cauchy horizon into a causality violating region, which would capture the notion of formation of a time machine. The strong cosmic censorship conjecture asserts that, for suitable initial data, the maximal development is generically inextendible. Hence, if correct, strong cosmic censorship excludes time machines because either they can't be formed or they are infinitely fine-tuned (non-generic).}

Let's consider now the argument of Ref.~\cite{Adams2006}, which discussed several flat spacetime field theories with superluminal propagation. The simplest example is a scalar field with action
\begin{equation}
 S =  -\frac{1}{2}  \int d^4 x \left[\eta^{\mu\nu} \partial_\mu\pi \partial_\nu \pi -\frac{c_3}{\Lambda^4} \left(  \eta^{\mu\nu} \partial_\mu\pi \partial_\nu \pi  \right)^2 \right],
  \end{equation}
where $\Lambda$ is a mass scale and $c_3$ a dimensionless constant. For this theory, the equation of motion is:
\begin{equation}
 E(\pi,\partial\pi,\partial^{2}\pi)\equiv \left(1-\frac{2c_{3}}{\Lambda^{4}}\partial\pi\cdot\partial\pi\right)\partial^{\mu}\partial_{\mu}\pi-\frac{4c_{3}}{\Lambda^{4}}\partial^{\mu}\pi\partial^{\nu}\pi\partial_{\mu}\partial_{\nu}\pi=0.
\end{equation}
For this equation, a hypersurface is characteristic if, and only if, it is null w.r.t. the (inverse) ``effective metric'':
\begin{equation}
G^{\mu\nu} (\partial \pi) = \left[1-\frac{2c_{3}}{\Lambda^{4}}(\partial\pi\cdot\partial\pi)\right] \eta^{\mu\nu} - \frac{4c_3}{\Lambda^4} \partial^\mu \pi \partial^\nu \pi.
\end{equation}
Assume that there exists a fiducial inertial frame for which $|\partial_\mu \pi| \ll \Lambda^2$.\footnote{This is probably required for the validity of effective field theory.} Then $G^{\mu\nu}$ has Lorentzian signature, which implies that the equation of motion is hyperbolic (a frame-independent statement).
It is $G^{\mu\nu}$ that determines causal properties of this equation. Following the terminology of the Introduction, we say that a covector $\xi$ is timelike iff it is timelike w.r.t. $G^{\mu\nu}$ etc.

Inverting $G^{\mu\nu}$ gives the effective metric. In the fiducial inertial frame we have
\begin{equation}
G_{\mu\nu}(\partial\pi) \approx \eta_{\mu\nu} + \frac{4c_3}{\Lambda^4} \partial_\mu \pi \partial_\nu \pi. 
\end{equation}
Contracting with a vector $X^\mu$ gives
\be
 G_{\mu\nu} X^\mu X^\nu \approx \eta_{\mu\nu} X^\mu X^\nu + \frac{4c_3}{\Lambda^4}  (X \cdot \partial \pi)^2.
\ee
From this it can be seen that the null cones of $G_{\mu\nu}$ and $\eta_{\mu\nu}$ are nested, with the null cone of $G_{\mu\nu}$ inside that of $\eta_{\mu\nu}$ when $c_3>0$ and outside that of $\eta_{\mu\nu}$ when $c_3<0$, i.e., the theory has superluminal propagation when $c_3<0$ \cite{Adams2006}.

We can now discuss the initial value problem. Given some inertial frame $x^\mu$, we would like to specify initial data $(\pi,\partial_0 \pi)$ on $\Sigma=\{x^0=0\}$. Of course $\Sigma$ is spacelike w.r.t. $\eta^{\mu\nu}$ but for a well-posed problem it is necessary that $\Sigma$ also be spacelike w.r.t. $G^{\mu\nu}$. For $c_3>0$ this is automatic. If $c_3<0$ then this appears to restrict our freedom to choose the initial data for $\partial_0 \pi$. But this is not a new restriction: we already imposed a restriction on the initial data, i.e., the existence of the fiducial inertial frame. In the fiducial frame, the surface $x^0=0$ is obviously spacelike w.r.t. $G^{\mu\nu}$. 

If, in some inertial frame, the surface $x^0=0$ is not spacelike w.r.t. $G^{\mu\nu}$ then the initial value problem will not be well-posed. In this case, for generic initial data, one would not expect a solution of the equation of motion to exist, even locally near $x^0=0$. One might be able to find a solution for very special initial data e.g. if the data is analytic then a solution will exist locally by the Cauchy-Kowalevskaya theorem. But this is infinitely fine-tuned: if one perturbs the initial data in a compact region then the resulting data will be non-analytic and no solution can be expected to exist. More generally, the solution does not depend continuously on the initial data. 

Ref. \cite{Adams2006} argues heuristically that it is possible to construct a time machine when $c_3<0$ by considering two lumps of scalar field, well-separated in the $x^2$ direction, which are highly boosted w.r.t. to each other in the $x^1$-direction. So consider initial data at $x^0=0$ consisting of two such lumps with a large relative boost.\footnote{
Note that such initial data will not satisfy the condition $|\partial_\mu \pi| \ll \Lambda^2$ everywhere.} The problem is that such an initial data surface is not everywhere spacelike w.r.t. $G^{\mu\nu}$ \cite{Adams2006} so this initial value problem is not well-posed. In general one would not expect any solution of the equations of motion to exist for such initial data. So one cannot build a time machine this way. 			

One might argue that it is obvious that a time machine could never result from Cauchy evolution of initial data since Cauchy evolution will break down when one is on the threshold of forming a time machine. So the question we should really ask is whether such a breakdown can occur starting from ``good'' initial data. In more physical terms: if one wishes to employ a large relative boost to build  a time machine one must specify how this large relative boost will arise from ``good'' initial data \cite{Geroch2010}. 

For the scalar field theory above, ``good'' means that the initial data surface $x^0=0$ should be spacelike w.r.t. $G^{\mu\nu}$. Cauchy evolution remains well-posed as long as surfaces of constant $x^0$ remain spacelike w.r.t. $G^{\mu\nu}$ so a necessary condition for formation of a time machine would be existence of a time $T>0$ at which the solution remains smooth but the surface $x^0=T$ becomes null w.r.t. $G^{\mu\nu}$ at one more more points. This would correspond to the threshold of formation of the time machine. 

Can this happen? It is well-known that such behaviour does not occur starting from {\it small} initial data, i.e., data such that $\pi$ and its first few derivatives are small. 
The solution arising from small initial data simply disperses in a similar way to a solution of the linear wave equation \cite{Christodoulou1986,Klainerman1986}. So superluminal propagation leads to no pathologies in the behaviour of solutions arising from small initial data. 
Not much is known about the global behaviour of solutions of nonlinear wave equations for {\it large} initial data. 
For most nonlinear equations, global regularity of solutions is not expected. Solutions can suffer shock formation, i.e., blow-up of the field $\pi$ (or a derivative of $\pi$) at some time $T>0$. See for example Ref. \cite{Miao2014} (albeit not for a Lorentz covariant equation). As far as we know, it is not excluded that the equation discussed above could have large data solutions that evolve to the threshold of formation of a time machine. But, as we have discussed, there is no compelling reason to believe that this is the case. 

We now turn to the proposal of Ref. \cite{Camanho2014} that it is possible to construct a time machine in EGB theory. This is done by exploiting the negative Shapiro time delay experienced by gravitons. The proposed time machine arises from two high energy gravitons, moving in opposite directions with non-zero impact parameter. Each graviton is described by an Aichelburg-Sexl ``shock-wave'' solutions \cite{Aichelburg1971,Dray1985}. It is assumed that the spacetime resulting from the collision is well-approximated by two outgoing Aicheburg-Sexl shock waves. Under these assumptions one can argue that there exist closed causal curves in the spacetime.

One problem with this construction is the use of Aichelburg-Sexl solutions. The curvature of an AS solution is a delta-function localized on a null hypersurface (with the amplitude of the delta-function diverging on a null line within this hypersurface: this is viewed as the worldline of the graviton). Owing to special symmetries of its curvature tensor, this is an exact solution of Einstein-Gauss-Bonnet theory. Now clearly one can superpose two such solutions moving in opposite directions (since the spacetime is flat between them) to obtain a solution valid until the two null hypersurfaces intersect. But  when these hypersurfaces intersect, it is far from clear that there is any sense in which the equation of motion can be satisfied. This is because the equation of motion involves products of curvature tensors. Hence along the line of intersection of the hypersurfaces, there will be a product of delta functions that cannot be balanced. Therefore it seems unlikely that the spacetime can be extended to the future of this intersection.\footnote{Note that a theory of interacting impulsive (i.e. delta-function curvature) gravitational waves does exist for GR \cite{Luk2012}.}

This problem arises from the fact that an AS solution is singular. So maybe we can solve the problem by smoothing out the singularity. As discussed in the Introduction, an AS solution can be obtained by taking a limit in which one boosts a black hole solution and takes the boost to infinity whilst scaling the black hole mass to zero, keeping the total energy fixed. This suggests that we should consider initial data consisting of two small (compared to the GB scale) black holes, moving with high relative boost in opposite directions with large impact parameter. It would be a difficult matter to construct such data explicitly, solving the constraint equations, but there is no reason to doubt that this can be done.

Now the question is whether this is ``good'' initial data. As discussed above, black holes with arbitrarily small mass don't exist for $d=5$. When $d=6$, small black holes are unphysical because the equation of motion is not hyperbolic. So consider $d \ge 7$. In section \ref{sec:speedlimit} we showed that there is a speed limit for small black holes arising from the condition that the initial data surface be spacelike. Hence we cannot start from initial data describing two black holes with a very large relative boost: such initial data will not be everywhere spacelike and hence this data cannot be evolved (or is infinitely fine-tuned), just as for the scalar field example discussed above. 

We can attempt to construct legitimate initial data by requiring that the speed limit is respected. Consider two small black holes, each of mass parameter \(\mu\), boosted in opposite directions with speed \(v\le v_{\rm max}\), separated in the transverse direction by a distance \(R\). Assume that the distance is sufficiently large for the gravitational interaction between the black holes to be negligible, i.e. \(R\gg L\). Consider a tensor-polarized graviton propagating between the holes. Our numerical results suggested that, for a black hole at rest, the maximum time advance $|D|$ that a graviton can experience is of order \(L\) and is achieved for \(b\sim L\). If we now boost the black hole in a direction transverse to the motion of the graviton (in order to avoid issues related to the length contraction effect), the time advance gets amplified by \(\gamma=(1-v^{2})^{-1/2}\): \(\abs{D}\sim \gamma L\). For the argument of \cite{Camanho2014} to work we need the time advance to `compensate' the time taken by the graviton to travel between the two black holes, i.e. \(\abs{D}\sim R\). However for this to hold we would need \(\gamma L \sim R \gg L\), that is \(\gamma\gg1\), which cannot be achieved because of the restriction $v < v_{\rm max}$.

In summary, we have argued that attempting to build a time machine spacetime in EGB theory using the method suggested in Ref. \cite{Camanho2014} will not work. This is because the initial data required is not everywhere spacelike (in the sense defined in the Introduction) so the initial value problem is not well-posed: either no solution will exist, or it will be infinitely fine-tuned.

\section*{Acknowledgements} 

This work was supported by ERC grant No. ERC-2011-StG 279363-HiDGR and by an STFC studentship. 



\appendix
\addtocontents{toc}{\protect\setcounter{tocdepth}{1}}

\section{Causal structure of a PDE}
\label{sec:causal_structure_appendix}
We give here a more rigorous treatment of the causal structure of a hyperbolic system, introducing the necessary concepts in PDE theory. We refer to \cite{Christodoulou2008,Choquet-Bruhat2008a,CourantHilbert,Evans1998} for the details. For simplicity we will limit our discussion to second order PDEs. Consider a second order linear differential operator \(L\) on a \(d\)-dimensional Lorentzian manifold \(\left(\mathcal{M},g\right)\), \(p\in\MM\). In local coordinates:
\begin{equation}
	\label{eq:diff_op}
	L=a^{\alpha\beta}(p)\partial_{\alpha}\partial_{\beta}+b^{\gamma}(p)\partial_{\gamma}+c(p)
\end{equation}

We then define the following:
\begin{definition}
	Let \(\xi\in T^{*}_{p}\MM\). The \emph{principal part} of the differential operator \(L\) is:
	\begin{equation}
	PP(L)=a^{\alpha\beta}(p)\partial_{\alpha}\partial_{\beta}.
	\end{equation}
	The \emph{principal symbol} of the operator is defined as the contraction:
	\begin{equation}
		P(p,\xi)=a^{\alpha\beta}(p)\xi_{\alpha}\xi_{\beta}.
	\end{equation}
\end{definition}
\noindent More generally, for a system of non-linear PDEs, \(u:\MM \rightarrow \reals^n\):
\begin{equation}
	\label{eq:nonlinpde}
	F_{I}(p,u,\partial u,\partial^{2}u)=0,\qquad 1 \le I \le n
\end{equation}
one defines the principal symbol as the principal symbol of the linearised system:
\begin{definition}
	The \emph{principal symbol} of \eqref{eq:nonlinpde} is:
	\begin{equation}
		P(p,\xi)_{IJ}=\frac{\partial F_{I}(p,u,\partial u,\partial^{2} u)}{\partial\left(\partial_{\alpha}\partial_{\beta}u^{J}\right)}\,\xi_{\alpha}\xi_{\beta}\,
	\end{equation}
\end{definition}

\begin{definition}
	Let \(\xi\in T^{*}_{p}\MM\). The \emph{characteristic polynomial} \(Q(p,\xi)\) is defined as:\footnote{
In a Lovelock theory (or in GR), this definition needs modification because of the gauge symmetry. See Ref. \cite{Reall2014} for details.}	
	\begin{equation}
		Q(p,\xi)=\det P(p,\xi).
	\end{equation}
\end{definition}
\begin{definition}
	We define the \emph{characteristic subset} of \(T^{*}_{p}\MM\) (or \emph{normal cone at \(p\)}) as:
	\begin{equation}
		\mathcal{C}^{*}_{p}=\{\xi\in T^{*}_{p}\MM \;:\; Q(p,\xi)=0 \}.
	\end{equation}
\end{definition}
\begin{remark}
	The \emph{characteristic polynomial} is a homogeneous polynomial in \(\xi\) of degree \(2n\).
\end{remark}
\noindent We can now define characteristic hypersurfaces.
\begin{definition}
	Consider a hypersurface \(\Sigma\subset\MM\) with normal 1-form \(\xi \). $\Sigma$ is a \emph{characteristic hypersurface} if \(Q(p,\xi)=0\), \(\forall p\in \Sigma\).
\end{definition}

Consider a hypersurface defined as a level set of some function \(\phi\). The surface will be characteristic iff \(Q(p,\dd \phi)=0\). This defines a first order non-linear equation for the function \(\phi\) know as the \emph{eikonal equation}.
\begin{remark}
	Such characteristic hypersurfaces are generated by \emph{bicharacteristic curves}. These are defined as the curves \((x(\lambda),\xi(\lambda))\) in the cotangent bundle \(T^{*}\MM\) satisfying:
	\begin{equation}
		\dot{x}^{\mu}=\frac{\partial Q(x,\xi)}{\partial \xi_{\mu}} \qquad \dot{\xi}_{\mu}=-\frac{\partial Q(x,\xi)}{\partial x^{\mu}}
	\end{equation}
	\noindent with the initial condition \(Q(p,\xi)=0\), which is preserved along these curves by construction.
\end{remark}
\noindent We now want to define the dual of the set \(\mathcal{C}_{p}^{*}\). 
\begin{definition}
	The \emph{ray cone} at \(p\in\MM\) is defined as the set
	\begin{equation}
		\mathcal{C}_{p}=\bigcup_{\xi\in \mathcal{C}^{*}_{p}}\mathcal{C}_{p,\xi}
	\end{equation}
	\noindent where 
	\begin{equation}
		\mathcal{C}_{p,\xi}=\{X \in T_{p}\MM \;:\; \langle X, \xi \rangle = 0\}
	\end{equation}
\end{definition}

\begin{remark}
	The projections of the bicharacteristics onto the base manifold are called \emph{rays} and their tangent vectors \(Y(p)=\frac{\partial Q(p,\xi)}{\partial \xi_{\mu}}\frac{\partial}{\partial x^{\mu}}\in \mathcal{C}_{p}\) are called \emph{characteristic tangent vectors}. The ray cone is nothing else that the `light cone' defined by null rays. 
\end{remark}
We now define what we mean by hyperbolic systems:
\begin{definition}
	The characteristic polynomial \(Q(p,\xi)\) is said to be \emph{hyperbolic} if there exists \(\zeta\in T^{*}_{p}\MM\) such that any straight line with direction \(\zeta\), such that it does not intersect the vertex \(\xi=0\), intersects the cone \(Q(p,\xi)=0\) in \(m n\) real roots, where \(m\) is the order of the PDE and \(n\) the rank of the principal symbol.

	A system of PDEs is \emph{hyperbolic} at \(p\in\MM\) if its characteristic polynomial is hyperbolic. We say that the system is \emph{strictly hyperbolic} if, moreover, all the roots are distinct.
\end{definition}

Denote by \(C_{p}^{*}\) the closure of the subset of \(T_{p}^{*}\MM\) defined by such one-forms \(\zeta\). It can be shown that \(C_{p}^{*}\) consists of two convex, opposite cones \(C_{p}^{*}=C_{p}^{*,+} \cup C_{p}^{*,-}\). These are non-empty and \(\partial C_{p}^{*}\subset \mathcal{C}_{p}^{*}\). Similarly, denote the dual cones in the tangent space by \(C_{p}=C_{p}^{+}\cup C_{p}^{-}\). In this case we have \(\mathcal{C}_{p}\subset C_{p}\). If it is possible to continuously distinguish between the convex cones \(C_{p}^{\pm}\), we say that the spacetime is \emph{time-orientable}. 
We can finally define the causal structure:
\begin{definition}
	A vector \(X\in T_{p}\MM\) is \emph{causal} if \(X\in C_{p}\).
	A one-form \(\xi\in T_{p}^{*}\MM\) is \emph{causal} if \(\xi \in C_{p}^{*}\).
\end{definition}
\noindent It follows, in particular, that a hypersurface \(\Sigma\) is \emph{spacelike} if its normal one-form is in the interior of \(C_{p}^{*}\), \(\forall p\in\Sigma\). Thus in discussing the initial value problem we shall prescribe initial data on a hypersurface which is spacelike according to this definition. 

\begin{example}
	Consider the class of solutions to Lovelock theory for which the characteristic polynomial factorises as a product of quadratic factors:
	\begin{equation}
		Q(p,\xi)=(G_{1}^{ab}(p)\xi_{a}\xi_{b})^{p_1}(G_{2}^{cd}(p)\xi_{c}\xi_{d})^{p_2}\ldots
	\end{equation}
	\noindent It is easy to see that the normal cone corresponds to the union of the null cones of the inverse effective metrics, while the ray cone corresponds to the union of the null cones of the effective metrics.
	The set \(C_{p}^{*}\) corresponds to the intersection of the subsets of the cotangent space defined by \(G_{I}^{ab}(p)\xi_{a}\xi_{b}\le 0\). The set \(C_{p}\), instead, corresponds to the union of the subsets of the tangent space for which \(G_{Iab}(p)X^{a}X^{b}\le 0\). In the case in which these cones form a nested set (e.g. for Ricci flat type N spacetimes) then \(C_{p}^{*}\) (\(C_{p}\)) corresponds to the innermost (outermost) cone in \(T_{p}^{*}\MM\) (\(T_{p}\MM\)). 
	In this case we see that a spacelike hypersurface is one which lies always outside of the outermost cone in the tangent space and whose normal vector always lies in the innermost cone in the cotangent space.	
\end{example}

\section{Perturbative calculations} 
\label{sec:approximation}
We give here more details on the perturbative calculation of the time delay and deflection angle. Recall from section \ref{sub:time_advance_in_egb} that we want to compute:
\begin{equation}
	\Delta \phi  = 2 b \int_{R_{0}}^{R} dr \; c_{A}(r)r^{-2}h(r),\quad
	\Delta t = 2\int_{R_{0}}^{R} dr  \; f(r)^{-1} h(r),
\end{equation}
\noindent where we have introduced:
\begin{equation}
	h(r)=\left(1-\frac{f(r)c_{A}(r)b^{2}}{r^2}\right)^{-1/2}.
\end{equation}
We want to calculate the above quantities subject to the assumption that $R_0$ is large compared to the black hole size in the following sense
\begin{equation}
	\label{eq:assumptions1}
	\frac{\mu}{R_{0}^{d-3}}\ll1, \quad \frac{L}{R_{0}}\ll 1
\end{equation}
We will assume that the cavity radius is large:
\be
\label{eq:assumptions1a}
\frac{R}{R_0} \gg 1
\ee
For the time delay we will need to assume that the cavity radius is not {\it too} large:
\begin{equation}
	\label{eq:assumptions2}
	 \frac{R}{R_{0}} \ll \frac{R_{0}^{d-3}}{\mu},\left(\frac{R_{0}}{L}\right)^{d-1}.
\end{equation}

\subsection{Approximation for $h(r)$}

The impact parameter is related to $R_0$ by equation (\ref{bR0}). Using \eqref{eq:assumptions1} we have:
\begin{equation}
	b^{2}=R_{0}^2\left(1+\frac{\mu}{R_{0}^{d-3}}\left(1-\frac{2\alpha\beta_{A}}{R_{0}^{2}}\right)\right)+\ldots
\end{equation}
\noindent similarly, introducing \(z=R_{0}/r \) (so $0<z \le 1$):
\begin{equation}
	\frac{f(r)c_{A}(r)}{r^2}=\frac{z^{2}}{R_{0}^{2}}\left(1-\frac{\mu}{R_{0}^{d-3}}\left(z^{d-3}-\frac{2\alpha\beta_{A}}{R_{0}^{2}}z^{d-1}\right)\right)+\ldots
\end{equation}
\noindent And hence:
\begin{equation}
	h(r)=(1-z^{2})^{-1/2}\left(1+\frac{1}{2}\frac{\mu}{R_{0}^{d-3}}\frac{z^2}{1-z^2}\left((1-z^{d-3})-\frac{2\alpha\beta_{A}}{R_{0}^{2}}(1-z^{d-1})\right)\right)+\ldots
\end{equation}
and the ellipsis denotes terms of order \({\cal O}\left(\left(\frac{\mu}{R_{0}^{d-3}}+\frac{L^{d-1}}{R_{0}^{d-1}}\right)^{2}\right)\).

\subsection{Approximation for the deflection angle}
Changing the integration variable to \(z\):
\begin{equation}
	\Delta \phi=2\frac{b}{R_{0}}\int_{R_{0}/R}^{1} dz \,c_{A}h
\end{equation}
In our approximation, denoting by \(\ldots\) terms of order \({\cal O}\left(\left(\frac{\mu}{R_{0}^{d-3}}+\frac{L^{d-1}}{R_{0}^{d-1}}\right)^{2}\right)\), we obtain: 
\begin{equation}
	c_{A}h=(1-z^{2})^{-1/2}\left(1+\frac{\mu}{R_{0}^{d-3}}\left(\frac{1}{2}z^{2}\frac{(1-z^{d-3})}{(1-z^{2})}-\frac{\alpha}{R_{0}^{2}}\beta_{A}\left(\frac{z^{2}+z^{d+1}-2z^{d-1}}{1-z^{2}}\right)\right)\right)+\ldots
\end{equation}
\noindent Which yields:
\begin{equation}
	\Delta\phi\,=2\arccos\left(\frac{R_{0}}{R}\right)+\frac{\mu}{R_{0}^{d-3}} \left[2\left(\frac{1}{2}-\frac{\alpha\beta_{A}}{R_{0}^{2}}\right)\arccos\left(\frac{R_{0}}{R}\right)+2J\right]+\ldots
\end{equation}
\noindent where we have defined:
\begin{equation}
	J=\int_{R_0/R}^{1} dz\, (1-z^{2})^{-1/2}\,\left(\frac{1}{2}z^{2}\frac{(1-z^{d-3})}{(1-z^{2})}-\frac{\alpha\beta_{A}}{R_{0}^{2}}\left(\frac{z^{2}+z^{d+1}-2z^{d-1}}{1-z^{2}}\right)\right).
\end{equation}
Using the large cavity radius assumption (\ref{eq:assumptions1a}) we have \(J=J_0+{\cal O}(R_{0}^{2}/R^{2})\), where
\begin{equation}
	J_0=\sqrt{\pi} (d-1)\left(1-\frac{2\alpha\beta_{A}}{R_{0}^{2}}\frac{(d-2)}{(d-1)}\right)\frac{\Gamma\left(\frac{d}{2}\right)}{2\Gamma\left(\frac{d+1}{2}\right)}-\frac{\pi}{2}\left(\frac{1}{2}-\frac{\alpha\beta_{A}}{R_{0}^{2}}\right).
\end{equation}
\noindent Moreover we have that
\begin{equation}
	2\arccos\left(\frac{R_{0}}{R}\right)=\pi-2\frac{R_{0}}{R}+{\cal O}\left(\frac{R_{0}^{2}}{R^{2}}\right)
\end{equation}
and hence
\begin{equation}
	2\left(\frac{1}{2}-\frac{\alpha\beta_{A}}{R_{0}^{2}}\right)\arccos\left(\frac{R_{0}}{R}\right)+2J=\sqrt{\pi} (d-1)\left(1-\frac{2\alpha\beta_{A}}{R_{0}^{2}}\frac{(d-2)}{(d-1)}\right)\frac{\Gamma\left(\frac{d}{2}\right)}{2\Gamma\left(\frac{d+1}{2}\right)}+{\cal O}\left(\frac{R_{0}}{R}\right).
\end{equation}
We can then conclude that
\begin{align}
	\nonumber \Delta\phi=&\:\pi-2\frac{R_{0}}{R}+\frac{\mu}{R_{0}^{d-3}} \left[\sqrt{\pi} (d-1)\left(1-\frac{2\alpha\beta_{A}}{R_{0}^{2}}\frac{(d-2)}{(d-1)}\right)\frac{\Gamma\left(\frac{d}{2}\right)}{2\Gamma\left(\frac{d+1}{2}\right)}\right]+\\
	&+{\cal O}\left(\left[\frac{R_0}{R}+\left(\frac{\mu}{R_{0}^{d-3}}+\frac{L^{d-1}}{R_{0}^{d-1}}\right)\right]^{2}\right).
\end{align}

\subsection{Approximation for the proper time}
Denoting again by \(\ldots\) terms of order \({\cal O}\left(\left(\frac{\mu}{R_{0}^{d-3}}+\frac{L^{d-1}}{R_{0}^{d-1}}\right)^{2}\right)\), we have:
\begin{equation}
	f(z)^{-1}h(z)=(1-z^{2})^{-1/2}\left(1+\frac{\mu}{R_{0}^{d-3}} z^{2}\left(z^{d-5}+\frac{1}{2}\frac{(1-z^{d-3})}{1-z^{2}}-\frac{2\alpha\beta_{A}}{R_{0}^{2}}\frac{(1-z^{d-1})}{1-z^{2}}\right)\right)+\ldots
\end{equation}
\noindent from which:
\begin{equation}
	\Delta t =2\sqrt{R^{2}-R_{0}^{2}}+2\frac{\mu}{R_{0}^{d-4}} I+\ldots
\end{equation}
\noindent where:
\begin{equation}
	I=\int_{R_0/R}^{1} dz\, (1-z^{2})^{-1/2}\,\left(z^{d-5}+\frac{1}{2}\frac{(1-z^{d-3})}{(1-z^{2})}-\frac{\alpha\beta_{A}}{R_{0}^{2}}\frac{(1-z^{d-1})}{1-z^{2}}\right).
\end{equation}
For large cavity radius, i.e., \eqref{eq:assumptions1a}, we have:
\begin{equation}
	\Delta t=2R-\frac{R_{0}^{2}}{R}+2\frac{\mu}{R_{0}^{d-4}}I_0+{\cal O}\left(R_{0}\left[\frac{R_0}{R}+\left(\frac{\mu}{R_{0}^{d-3}}+\frac{L^{d-1}}{R_{0}^{d-1}}\right)\right]^{2}\right),
\end{equation}
where
\begin{align}
	I_0&=\sqrt{\pi}\left(1-\frac{2\alpha\beta_{A}}{R_{0}^{2}}\frac{(d-4)}{(d-3)}\right)\frac{(d-1)(d-3)}{(d-4)}\frac{\Gamma\left(\frac{d}{2}\right)}{4\Gamma\left(\frac{d+1}{2}\right)}.
\end{align}
Plugging back in the above we obtain
\begin{align}
	\nonumber \Delta t = &\: 2R + \frac{\mu}{R_{0}^{d-4}}\left[\sqrt{\pi}\left(1-\frac{2\alpha\beta_{A}}{R_{0}^{2}}\frac{(d-4)}{(d-3)}\right)\frac{(d-1)(d-3)}{(d-4)}\frac{\Gamma\left(\frac{d}{2}\right)}{2\Gamma\left(\frac{d+1}{2}\right)}\right]+\\
	&+{\cal O}\left(R_{0}\left[\frac{R_0}{R}+\left(\frac{\mu}{R_{0}^{d-3}}+\frac{L^{d-1}}{R_{0}^{d-1}}\right)\right]^{2}+\frac{R_{0}^{2}}{R}\right).
\end{align}
Finally, note that in this approximation we have \(f(R)=1+{\cal O}(\frac{\mu}{R^{d-3}})\) and $\mu/R^{d-3} =(\mu/R_0^{d-3})(R_0/R)^{d-3}$ which is negligible to the order of approximation used above, hence \(\Delta \tau \approx \Delta t \).

\subsection{Approximation for the time in Minkowski}
We have:
\begin{align}
	\nonumber 2 R \sin\left(\Delta\phi/2\right)= & \: 2R \sin\left[\arccos(R_{0}/R)+\frac{\mu}{R_{0}^{d-3}}\left(\left(\frac{1}{2}-\frac{\alpha\beta_{A}}{R_{0}^{2}}\right)\arccos\left(R_{0}/R\right)+J\right)+\ldots\right]\\
	\nonumber =& \: 2\sqrt{R^{2}-R_{0}^{2}}+\frac{\mu}{R_{0}^{d-4}}\left(\left(\frac{1}{2}-\frac{\alpha\beta_{A}}{R_{0}^{2}}\right)\arccos\left(R_{0}/R\right)+J\right)+\\
	& +{\cal O}\left( R\left( \frac{ \mu}{R_0^{d-3}} + \frac{ L^{d-1}}{R_0^{d-1}}\right)^2 \right).
\end{align}
We now use the condition \eqref{eq:assumptions2} that the cavity is not too large. This ensures that the last term above is small and we obtain:
\begin{align}
	\Delta t_{\rm Mink}=&2R +\frac{\mu}{R_{0}^{d-4}}\left(\sqrt{\pi}(d-1)\left(1-\frac{2\alpha\beta_{A}}{R_{0}^{2}}\frac{(d-2)}{(d-1)}\right)\frac{\Gamma\left(\frac{d}{2}\right)}{2\Gamma\left(\frac{d+1}{2}\right)}\right)+{\cal O}\left(R \left[\frac{R_{0}}{R}+\left( \frac{ \mu}{R_0^{d-3}} + \frac{L^{d-1}}{R_0^{d-1}}\right)\right]^2 \right).
\end{align}

\subsection{Approximation for the time delay}
Finally, putting the above results together, we have determined the time delay under the conditions  \eqref{eq:assumptions1}, \eqref{eq:assumptions1a}, \eqref{eq:assumptions2}:
\begin{equation}
	D=\frac{\mu}{R_{0}^{d-4}}\left[\sqrt{\pi} \left(1-\frac{2\alpha \beta_{A}}{R_{0}^{2}}\frac{(d-4)}{(d-1)}\right)\frac{(d-1)}{(d-4)}\frac{\Gamma\left(\frac{d}{2}\right)}{2\Gamma\left(\frac{d+1}{2}\right)}\right]+{\cal O}\left(R \left[\frac{R_{0}}{R}+\left( \frac{ \mu}{R_0^{d-3}} + \frac{L^{d-1}}{R_0^{d-1}}\right)\right]^2 \right).
\end{equation}

\bibliographystyle{JHEP}
\bibliography{time_delay}
\end{document}